\documentclass[aps,prc,superscriptaddress,showpacs,nofootinbib,twocolumn]{
revtex4}

\usepackage{graphicx,amssymb,amsmath,amsfonts,dcolumn,gensymb,hyperref}
\usepackage[dvipsnames]{xcolor}
\usepackage[normalem]{ulem}

\allowdisplaybreaks[1]

\newcolumntype{d}[1]{D{.}{.}{#1}}

\newcommand{\unit}[1]{\ensuremath{\,\mathrm{#1}}}

\newcommand{\etal}{\textit{et al.} }
\newcommand{\etaln}{\textit{et al.}}

\newcommand{\tr}{{\rm tr}}
\newcommand{\rbaryon}{{\rm baryon}}
\newcommand{\rsat}{{\rm sat}}
\newcommand{\rsym}{{\rm sym}}
\newcommand{\rkin}{{\rm kin}}
\newcommand{\rpot}{{\rm pot}}
\newcommand{\rcrit}{{\rm crit}}
\newcommand{\rsnm}{{\rm SNM}}
\newcommand{\rpnm}{{\rm PNM}}
\newcommand{\ie}{\textit{i.e.}, }

\newcommand{\apjs}{Astrophys. J. Suppl. Ser.}

\usepackage{libertineotf}
\begin{document}

\title{Equation of state effects in the core collapse of a $20$-$M_\odot$ star}

\author{A. S. Schneider}\email{andre.schneider@astro.su.se}
\affiliation{Department of Astronomy and the Oskar Klein Centre, Stockholm
University, AlbaNova, SE-106 91 Stockholm, Sweden}
\affiliation{TAPIR, Walter Burke Institute for Theoretical Physics, MC 350-17,
California Institute of Technology, Pasadena, CA 91125, USA}

\author{L. F. Roberts}\email{robertsl@nscl.msu.edu}
\affiliation{National Superconducting Cyclotron Laboratory and Department of
Physics and Astronomy, Michigan State University, East Lansing, MI 48824, USA}

\author{C. D. Ott}\email{cott@ocslabs.com}
\affiliation{OCS Labs LLC, Pasadena, CA 91104}

\author{E. O'Connor}\email{evan.oconnor@astro.su.se}
\affiliation{Department of Astronomy and the Oskar Klein Centre, Stockholm
University, AlbaNova, SE-106 91 Stockholm, Sweden}
\date{\today}

\begin{abstract}

Uncertainties in our knowledge of the properties of dense matter near and above
nuclear saturation density are among the main sources of variations in
multi-messenger signatures predicted for core-collapse supernovae (CCSNe) and
the properties of neutron stars (NSs). 
We construct 97 new finite-temperature equations of state (EOSs) of dense matter 
that obey current experimental, observational, and theoretical 
constraints and discuss how systematic variations in the EOS parameters affect 
the properties of cold nonrotating NSs and the core collapse of a 
$20$-$M_\odot$ progenitor star.
The core collapse of the $20$-$M_\odot$ progenitor star is simulated in spherical
symmetry using the general-relativistic radiation-hydrodynamics code \texttt{GR1D}
where neutrino interactions are computed for each EOS using the \texttt{NuLib} 
library.
We conclude that the effective mass of nucleons at densities above nuclear
saturation density is the largest source of uncertainty in the CCSN neutrino 
signal and dynamics even though it plays a subdominant role in most properties 
of cold NS matter. 
Meanwhile, changes in other observables affect the properties of cold NSs,
while having little effect in CCSNe. 
To strengthen our conclusions, we perform six octant three-dimensional
CCSN simulations 
varying the effective mass of nucleons at nuclear saturation density.
We conclude that neutrino heating and, thus, the likelihood of explosion is
significantly increased for EOSs where the effective mass of nucleons at
nuclear saturation density is large. 

\end{abstract}


\pacs{21.65.Mn,26.50.+x,26.60.Kp}

\maketitle


\section{Introduction}\label{sec:Intro}

Stars with masses above roughly eight times the mass of the Sun ($M_\odot$), end their lives in a core collapse event, in many cases leading to a core-collapse supernova (CCSN) explosion. 
Core collapse sets in once electron degeneracy pressure in the nickel-iron core of a massive star can no longer support it against gravity \cite{woosley:02}.

Core collapse proceeds until the inner core reaches nuclear saturation density, $\rho_\rsat\gtrsim 2.7\times10^{14} \unit{g\,cm}^{-3}$, at a temperature of $10-20\unit{MeV}$. 
At this point, the residual nuclear force prevents the inner core from contracting any further and it rebounds into the still infalling outer core, creating a shock wave. 
As the shock wave propagates through the outer core it eventually stalls because of energy losses resulting from dissociation of heavy nuclei and to lesser extent due to neutrino losses from behind the shock.

A few mechanisms that revive the shock and lead to successful CCSNe have been suggested, see discussion in Ref.~\cite{burrows:13} and references therein. 
Simulations have shown that it is likely that a multitude of macroscopic (e.g., progenitor structure, large-scale convection, magnetohydrodynamic forcing) and microscopic properties and processes (e.g., neutrino heating) couple non-linearly to drive an explosion. Still, it is believed that the main contributor to shock revival is the neutrino heating mechanism \cite{wilson:85, bethe:85}, whereby $\sim10\%$ of the outgoing electron-flavor neutrino luminosity is deposited behind the shock.  
This provides the shock with thermal support, drives turbulence, and aids in shock runaway \cite{janka:01, couch:15, burrows:13}.

One of the fundamental ingredients to understand the dynamics of core collapse events is the equation of state (EOS) of dense matter.  
The density at which the collapse halts, how many protons are converted into neutrons during the collapse, the spectra of neutrinos, how much energy is deposited behind the shock and its expansion rate, the ejecta mass and its composition, the proto neutron star (PNS) mass, its radius, cooling rate, and whether it later collapses into a black-hole (BH) as well as the gravitational wave (GW) signal, are all dependent on the EOS. 
In a CCSN, and also in NS mergers, matter exists in a wide range of temperatures, $0\lesssim{T}\lesssim\mathcal{O}(100\unit{MeV})$, densities, ${\rho}\lesssim 10^{15}\unit{g\,cm}^{-3}$, and proton fractions, $0.0\lesssim{y}\lesssim0.5$.  Some of these conditions are so extreme they are not readily available to laboratory experiments and, thus, such regions of parameter space can only be probed indirectly from observations in consent with computational and theoretical models.

Recently Ref.~\cite{margueron:18a} introduced the concept of meta-modeling for the nuclear EOS (see also Ref.~\cite{steiner:05}). 
In their model, the EOS is parametrized in terms of empirical parameters, \ie nuclear matter binding energy, saturation density, incompressibility, symmetry energy, and so on.
The average values of the empirical parameters and their uncertainties are estimated based on experimental and theoretical nuclear physics constraints. 
In follow-up studies meta-modeling was used to study the effects of uncertainties in the empirical parameters on NS properties \cite{margueron:18b}, finite size effects in the description of nuclear masses and radii of ground state nuclei \cite{chatterjee:17}, and to compute correlations between empirical parameters from known constraints \cite{margueron:18c}.

We follow the meta-modeling approach \cite{margueron:18a, margueron:18b} and analyze how uncertainties in properties of nuclear matter affect cold NS properties and the core collapse of a $20$-$M_\odot$ progenitor star. 
We use the meta-modeling formalism to construct a family of finite temperature EOSs of dense matter.  
The EOSs are built using the recently developed open-source \texttt{SROEOS} code \cite{schneider:17}, which is itself based on the Lattimer and Swesty liquid-drop model of nuclei \cite{lattimer:91}, with a few improvements.  
The main improvements relevant to this work are the possibility to compute EOSs where (1) the effective mass of nucleons is different from their vacuum values and (2) for any desired value of the incompressibility of nuclear matter $K_\rsat$, instead of the canonical values of $180\unit{MeV}$, $220\unit{MeV}$, and $375\unit{MeV}$ to which the code of \cite{lattimer:91} is essentially limited. 
The \texttt{SROEOS} model has also been extended to transition to a description of many nuclear species in nuclear statistical equilibrium (NSE) at low densities.

The main goal of this study is to separately determine how each empirical parameter of the EOS may affect a core collapse event and the resulting PNS.
This is only possible using many EOSs obtained within a single formalism.
Previous studies have studied the effect of the EOS on CCSNe and their observables, e.~g., Refs.~\cite{oconnor:11, hempel:12, steiner:13, couch:13a, suwa:13, char:15, fischer:14, furusawa:17a, richers:17a, nagakura:18, morozova:18}.
The main drawback of these studies is that often the EOSs being compared were obtained with distinct approaches, used different prescriptions to describe low density matter, and, with the exception of Ref.~\cite{richers:17a} which analyzed changes resulting from using 18 different EOSs in their simulations, the number of EOSs investigated was rather small. 
Thus, in many cases, it was challenging to disentangle how a parameter of the EOS contributed to a given observable.


In this paper, we focus on EOS effects on the neutrino heating mechanism and delay the study of GW signals to future work.  
We simulate the core collapse of a single non-rotating $20$-$M_\odot$ progenitor star taken from \cite{woosley:07} using 97 distinct EOSs that each vary in at most two different empirical parameters from a baseline EOS.  
The \texttt{SROEOS} code is ideal for this type of sensitivity study as it allows one to compute many EOSs within the same framework using arbitrary Skyrme-type parametrizations of the nuclear forces.  
Furthermore, to limit our assessment only to the effects of the high-density
part of the EOS, we use the same nuclear surface parametrization for all EOSs and the same NSE EOS at low densities for all simulations.  
The CCSN simulations are performed using the open-source general-relativistic multi-group radiation-hydrodynamics code \texttt{GR1D} \cite{oconnor:10, oconnor:11} and the \texttt{NuLib} neutrino transport libraries \cite{oconnor:13, oconnor:15}.  
Since the \texttt{GR1D} code is limited to spherical symmetry, we also perform six three-dimensional (3D) simulations, limited to an octant of the 3D cube to keep computational demands manageable. 
For this, we employ the open-source 3D general-relativistic radiation-hydrodynamics code \texttt{Zelmani} \cite{roberts:16,ott:18}, which is based on the \texttt{Einstein Toolkit} \cite{loeffler:12, moesta:14a}.  
We perform the octant 3D runs for five variations of the SLy4 EOS \cite{chabanat:98} and the LS220\footnote{LS220 is the Lattimer \& Swesty EOS with incompressibility $K_\rsat=220\unit{MeV}$ EOS. 
In this work the LS220 EOS was recomputed using the \texttt{SROEOS} code \cite{schneider:17}.} EOS
\cite{lattimer:91}.

This paper is structured as follows.  
In Section \ref{sec:Meta_EOS}, we discuss a variant of the meta EOS model of Ref.~\cite{margueron:18a} that suits our needs.  
We proceed to discuss how each of the empirical parameters affects the properties of cold beta-equilibrated NSs in Sec.~\ref{sec:NSs} and spherically-symmetric core collapse in Sec.~\ref{sec:CCSN}.  
In Sec.~\ref{sec:Octant}, we discuss 3D runs with octant symmetry. 
We conclude in Sec.~\ref{sec:Conclusions}.

\section{Meta EOS}
\label{sec:Meta_EOS}

Motivated by Ref.~\cite{margueron:18a}, we use a metamodeling formalism to compute Skyrme parameters for the nucleonic EOS in terms of empirical nuclear parameters.  
In this work, matter is assumed to be made solely of nucleons, electrons, positrons, and photons\footnote{It is expected that the EOS softens at very high temperatures and densities due to the appearance of heavy   leptons, hyperons, condensates, and quark-gluon plasmas \cite{sagert:09}. 
They are not explicitly included here since we take a parameterized approach to the high-density EOS. }.
Electrons, positrons, and photons are treated as uniform free gases and charge neutrality is assumed.  
Therefore, their contributions to the EOS decouple from the nucleon contributions. 
Our treatment of these components of the EOS is discussed in detail in Appendix A of Ref. \cite{schneider:17}.

\subsection{Skyrme Model}

The bulk nuclear contribution to the EOS is computed assuming non-relativistic effective Skyrme-type nucleon-nucleon interactions.
In this approach the energy per baryon $\epsilon_B$ of nucleonic matter with number density $n$ and proton fraction $y$ can be separated into its kinetic and potential energy density contributions \textit{i.e.},
\begin{equation}\label{eq:Skyrme}
 \epsilon_B(n,y,T) = \epsilon_\rkin(n,y,T) + \epsilon_\rpot(n,y)\,.
\end{equation}

The kinetic energy density term is 
\begin{equation}\label{eq:ekin}
\epsilon_\rkin(n,y,T)=\frac{1}{n}\left(\frac{\hbar^2\tau_n}{2m_n^\star}
+\frac{\hbar^2\tau_p}{2m_p^\star}\right)\,\,,
\end{equation}
where
\begin{equation}\label{eq:tauT}
 \tau_t=\frac{1}{2\pi^2}\left(\frac{2m_t^\star T}{\hbar^2}\right)^{\frac{5}{2}}
\mathcal{F}_{3/2}(\eta_t)\,,
\end{equation}
and the density dependent effective nucleon masses $m_t^\star$ are given by
\begin{equation}\label{eq:mstar}
 \frac{\hbar^2}{2m_t^\star}=
\frac{\hbar^2}{2m_t}+\alpha_1 n_{t}+\alpha_2n_{-t}\,.
\end{equation}
Above, $n_t$ and $m_t$ are, respectively, the density and vacuum mass of a nucleon with isospin $t$, where $t=n$ for neutrons and $t=p$ for protons, and, if $t=n$ then $-t=p$ and vice-versa.
The neutron and proton densities are related to the proton fraction $y$ and the nucleon density $n$ by $n_n=(1-y)n$ and $n_p=yn$, respectively. 
The quantities $\alpha_1$ and $\alpha_2$ are parameters of the model and establish a simple dependence of the nucleon effective masses on the density and proton fraction of the system. 
We stress that the Skyrme model treatment of effective masses is rudimentary, other models allow for much more complex dependencies of $m^\star$ \cite{constantinou:15b}. 
Nevertheless, we use this model as a guide to teach us how each piece of the EOS affects neutron star (NS) properties and the dynamics of CCSNe.

The Fermi integral in Eq. \eqref{eq:tauT} is defined as
\begin{equation}\label{eq:Fermi}
\mathcal{F}_k(\eta)=\int_0^\infty\frac{u^k du}{1+\exp(u-\eta)} \,,
\end{equation} 
and is a function of the degeneracy parameter 
\begin{equation}\label{eq:eta}
 \eta_t=\frac{\mu_t-V_t}{T}\,.
\end{equation}
Here, $\mu_t$ is the nucleon chemical potential and $V_t$ is the single-particle potential (see \cite{schneider:17} for more details).

The temperature-independent potential energy density term in Eq. \eqref{eq:Skyrme} has the form
\begin{equation}\label{eq:epot}
 \epsilon_\rpot(n,y)=\sum_{i=0}^{N}
 \left[a_i+4b_iy(1-y)\right]n^{\delta_i}.
\end{equation}
where $a_i$, $b_i$, and $\delta_i$ are constant parameters of the Skyrme model.
The $i=0$ term is chosen to represent two-body nucleon interactions. 
Therefore, we fix $\delta_0=1$ for all models.
Meanwhile, the $i>0$ terms approximate effects of many-body interactions \cite{lattimer:91}.
The summation in most Skyrme parameterizations ends at $N=1$, while only a small number of studies in the literature consider $N>1$ \cite{dutra:12}. 
To allow for more flexibility in our empirically fitted models, we choose to fix $N=3$ and $\delta_0=1$, $\delta_1=4/3$, $\delta_2=2$, and $\delta_3=7/3$ (the last three terms amount to an expansion in terms of the Fermi momenta of the nucleons $k_t \propto n_t^{2/3}$ \cite{lim:17, tews:17}). 
Therefore, the EOS model contains ten free parameters $\{a_0, b_0, a_1, b_1, a_2, b_2, a_3, b_3, \alpha_1, \alpha_2\}$ that we fit using a set of empirical properties of nuclear matter.

\subsection{Empirical Parameters}

Now, we would like to define a set of empirical properties with which to constrain our Skyrme EOS parameters.  
First, we consider measurable properties of nearly symmetric nuclear matter near nuclear saturation density.  
In these conditions, the zero-temperature nuclear EOS can be expanded about nuclear saturation density, $n=n_{\rm sat}\simeq0.155\unit{fm}^{-3}$, for symmetric matter ($y=1/2$) in a Taylor series, giving rise to a set of expansion parameters that can be empirically constrained. 
This expansion is written as
\begin{equation}\label{eq:expansion}
 \epsilon_B(n,y)=\epsilon_{\rm is}(x)+\delta^2\epsilon_{\rm iv}(x)\,,
\end{equation}
where $x=(n-n_\rsat)/(3n_\rsat)$ and $\delta=1-2y$ is the isospin asymmetry.
Here, the isoscalar (is) and isovector (iv) expansion terms are \cite{margueron:18a, piekarewicz:09}
\begin{align}
\label{eq:is}
 \epsilon_{\rm is}(x) &=\epsilon_\rsat+\frac{1}{2!}K_\rsat x^2
          +\frac{1}{3!}Q_\rsat x^3+\hdots\,,\\
\label{eq:iv}
 \epsilon_{\rm iv}(x) &=\epsilon_\rsym+L_\rsym x+\frac{1}{2!}K_\rsym x^2
\nonumber\\ &\quad+\frac{1}{3!}Q_\rsym x^3+\hdots\,,
\end{align}
shown here explicitly up to third order. 
The empirical parameter $\epsilon_\rsat$ is the energy per baryon at nuclear saturation density  $n_\rsat$, $K_\rsat$ is the isoscalar incompressibility modulus, and $Q_\rsat$ the isoscalar skewness. 
Similarly, $\epsilon_\rsym$ is the symmetry energy, $L_\rsym$ is related to the slope of symmetry energy in the direction of increasing density, $K_\rsym$ is the isovector incompressibility modulus, and $Q_\rsym$ is the isovector skewness. 
By definition of the saturation density $n_\rsat$, the linear term in $x$ of $\epsilon_{\rm is}$ vanishes. 
In principle, all of these expansion parameters can be determined experimentally, with varying degrees of difficulty. 
Nevertheless, the lower-order parameters are substantially easier to constrain. 
Therefore, we only include the well constrained saturation density empirical parameters $\{n_\text{sat}, \epsilon_\text{sat}, K_\text{sat}, \epsilon_\text{sym}, L_\text{sym}, K_\text{sym}\}$ in our Skyrme model fits described below.

Although this expansion is useful near saturation density, it cannot accurately describe the behavior of the nuclear EOS at densities larger than a few times saturation density since $x$ is no longer small and the expansion breaks down. 
Densities this large are reached in CCSNe and in the cores of NSs. 
Therefore, we also require empirical constraints at higher density. 
Most experiments probe densities near saturation density, but there are some results available for higher densities. 
Using measurements of flow in heavy ion collisions and theoretical transport models, \cite{danielewicz:02} constrained the baryonic pressure $P_B=n^2\partial\epsilon_B/\partial n$ of symmetric nuclear matter (SNM) and pure neutron matter (PNM), albeit in a model dependent way, at four times nuclear saturation density, $P_\text{SNM}^{(4)}=P_B(n=4 n_\text{sat}, y=1/2)$ and $P_\text{PNM}^{(4)}=P_B(n=4 n_\text{sat}, y=0)$.
Constraints on these pressures have recently been made sharper by combining the results of these flow experiments with constraints on the tidal deformability of NSs inferred from GW170817 \cite{tsang:18}.

Finally, although they do not enter into the expansion above, the nucleon effective masses at saturation density can also be considered a quasi-empirical parameter \cite{margueron:18a}. 
However, there is considerable complexity involved in extracting this property of the single quasi-particle energies.  
Nevertheless, the nucleon effective masses are particularly important for determining the temperature dependence of the nuclear EOS [see Eq.~\eqref{eq:ekin} above]. 
Therefore, we include the nucleon effective mass at saturation density in SNM, $m^\star \equiv m_n^\star(n=n_\rsat, y=1/2)$, and the neutron-proton effective mass splitting in PNM, $\Delta m^\star \equiv m_n^\star(n=n_\rsat, y=0) - m_p^\star(n=n_\rsat, y=0)$, in our list of empirical parameters. 

In total, this gives ten empirical parameters that we consider in this work, $\{n_\text{sat}, \epsilon_\text{sat}, K_\text{sat}, \epsilon_\text{sym}, L_\text{sym}, K_\text{sym}, m^\star, \Delta m^\star, P_\text{PNM}^{(4)}, P_\text{SNM}^{(4)}\}$.  
Due to their small uncertainties, we fix the values of the nuclear saturation number density $n_\rsat=0.155\unit{fm}^{-3}$ (mass density $\rho_\rsat = 2.7 \times 10^{14}\unit{g\,cm}^{-3}$) and of the energy at nuclear saturation density $\epsilon_\rsat=-15.8\unit{MeV}$.  
Other saturation density quantities are allowed to vary within their experimental or theoretical uncertainties (as compiled in \cite{margueron:18a}) as long as they are able to produce $2$-$M_\odot$ NSs \cite{demorest:10,antoniadis:13,fonseca:16}.  
The exception to this choice is the slope of the symmetry energy $L_\rsym$.  
Instead of using the average values of Ref.~\cite{margueron:18a}, $L_\rsym = 60 \pm 15 \unit{MeV}$, we set $L_\rsym = 45 \pm 7.5 \unit{MeV}$.
Although this choice only probes the lower half of possible values compiled in Ref.~\cite{margueron:18a}, we choose these limits so that the mass-radius relationships of NSs in this work are centered near the center of the constraints computed from observations of x-ray bursts \cite{nattila:16}. 
These limits also agree with combined theoretical calculations of pure neutron matter and astrophysical observations \cite{lattimer:13a, tews:17, oertel:17}.
Even though $L_\rsym$ is correlated with radii of low mass NSs \cite{alam:16}, for the systems we study, our limited choice for $L_\rsym$ has little effect on PNS properties in the first second after core collapse.  
Finally, we ignore existing correlations between the different empirical nuclear matter parameters \cite{alam:16, tews:17, margueron:18c}.  
Note, however, that the allowed ranges for empirical parameters contain EOSs that do not fulfill expected correlation between $\epsilon_\rsym$ and $L_\rsym$ determined on the basis of unitary gas considerations \cite{tews:17}.
We justify our choice with our primary interest in how different parameters of the EOS affect CCSNe. 
Our focus is less on particularly intricate details of the EOS.  
In Tab. \ref{tab:constraints} we summarize the constraints used in this work.

\begin{table}[htbp]
\caption{\label{tab:constraints} 
Constraints of nuclear matter properties used in this work grouped in sets defined in Sec.~\ref{ssec:sets}. 
Nuclear matter empirical parameters were compiled in Ref.~\cite{margueron:18a}, see references therein for details. 
Meanwhile, nuclear matter pressure at $4n_\rsat$, $P^{(4)}$, for SNM and PNM is from Ref.~\cite{danielewicz:02}.  
We use values similar to the ones in Refs.~\cite{margueron:18a, danielewicz:02}, but exclude from our analysis regions of parameter space that fail to reproduce $2$-$M_\odot$ NSs and, in the case of $L_\rsym$, values that lead to too large radii for NSs \cite{nattila:16}. 
We show the averages and one-standard deviations compiled or assumed in this work.}
\begin{ruledtabular}
\begin{tabular}
{c  c  D{,}{\pm}{-1}  D{,}{\pm}{-1}  l}
\multicolumn{1}{c}{Set}&
\multicolumn{1}{c}{Quantity}&
\multicolumn{1}{c}{Range} &
\multicolumn{1}{c}{This work} &
\multicolumn{1}{c}{Units} \\
\hline
$s_M$&$m^\star$
& 0.75 , 0.10  & 0.75 , 0.10 & $m_n$ \\
&$\Delta m^\star$
& 0.10 , 0.10  & 0.10 , 0.10 & $m_n$ \\
\hline
$-$&$n_\rsat$
& 0.155 , 0.005 & 0.155  & $\unit{fm}^{-3}$\\
&$\epsilon_\rsat$
& -15.8 , 0.3  & -15.8   &
\footnotesize{$\unit{MeV\,baryon}^{-1}$}\\
\hline
$s_S$&$\epsilon_\rsym$
& 32 , 2         & 32 , 2     &
\footnotesize{$\unit{MeV\,baryon}^{-1}$}\\
& $L_\rsym$
& 60 , 15        & 45 , 7.5   &
\footnotesize{$\unit{MeV\,baryon}^{-1}$}\\
\hline
$s_K$& $K_\rsat$
& 230 , 20       & 230 , 15   &
\footnotesize{$\unit{MeV\,baryon}^{-1}$}\\
& $K_\rsym$
& -100 , 100     & -100 , 100 &
\footnotesize{$\unit{MeV\,baryon}^{-1}$}\\
\hline
$s_P$ & $P^{(4)}_{\rsnm}$
& 100 , 50 & 125 , 12.5 & {$\unit{MeV\,fm}^{-3}$} \\
&
$P^{(4)}_{\rpnm}$
& 160 , 80 & 200 , 20   & {$\unit{MeV\,fm}^{-3}$}\\
\end{tabular}
\end{ruledtabular}
\end{table}

\subsection{Empirically Constrained Skyrme EOS Models}
\label{ssec:sets}

For a given set of Skyrme parameters, the empirical parameters described in the last section can be calculated from the Skyrme energy density [Eq.~\eqref{eq:Skyrme}], its derivatives, and the Skyrme expression for the effective masses [Eq.~\eqref{eq:mstar}]. 
Conversely, for a given choice of the ten empirical parameters given above, the ten Skyrme parameters are fixed. Our method for finding the Skyrme
parameters from the empirical parameters is given in Appendix \ref{app:linsys}.  
We stress that the fitted Skyrme parameterization only matches the saturation density expansion [Eq.~\eqref{eq:expansion}] at saturation density since the Skyrme model has a different functional form from the polynomial expansion.

To investigate the impact of EOS uncertainties on cold NSs and core collapse, we build a set of 97 Skyrme EOSs by picking 97 sets of the empirical parameters in the ranges given in
Tab.~\ref{tab:constraints}.  
We initially set the quantities used to obtain the Skyrme parametrization to their average values.  
Then, two-sigma variations in the nuclear properties are implemented for four sets of nuclear properties with two quantities each.  
The sets are
\begin{subequations}\label{eq:sets}
\begin{align}
s_M&=\{m^\star, \Delta m^\star\}\,,\label{eq:sm}\\
s_S&=\{\epsilon_\rsym, L_\rsym\}\,, \label{eq:ss}\\
s_K&=\{K_\rsat, K_\rsym\}\,, \label{eq:sk}\\
s_P&=\{P^{(4)}_{\rsnm},P^{(4)}_{\rpnm}\} \label{eq:sp}\,.
\end{align}
\end{subequations}
Thus, for set $s_M$ the values of $m^\star$ and $\Delta m^\star$ can be their average values ($m^\star=0.75$ and $\Delta m^\star = 0.10$), or their average values plus or minus one standard deviation ($m^\star = 0.75 \pm 0.10$ and $\Delta m^\star = 0.10 \pm 0.10$) or two standard deviations ($m^\star = 0.75 \pm 0.20$ and $\Delta m^\star = 0.10 \pm 0.20$).
Similar variations are implemented for all other sets, leading to a total of 97 different parametrization for the EOS\footnote{There are 25 EOSs in each set $s$. However, the baseline EOS with the average values of the observables is the same for all 4 sets.}.
For each of the parametrizations we build an EOS table using the open-source \texttt{SROEOS} code we have recently developed \cite{schneider:17}.

\subsection{Non-uniform and low density matter}
\label{ssec:nonuniform}

To limit our focus to the effects of the empirical parameters on CCSNe, we set the same parametrization of the nuclear surface for all EOSs. 
This is different from what we presented in Ref.~\cite{schneider:17}, where the parametrization of the surface properties was computed self-consistently based on the Skyrme parameters.
We defer to future work a detailed study of nuclear surface effects on CCSNe. 
Here, the surface parameters are chosen to be $\sigma_s = 1.15 \unit{MeV\,fm}^{-2}$, $q=16$, $\lambda=3.0$, and $p=1.5$, see Eqs. (19) and (20) in Ref.~\cite{schneider:17}.
The surface parametrization chosen here leads to a surface symmetry energy $S_S = 57.8 \unit{MeV}$, in agreement with the value $S_S=58.9\pm1.1\,\mathrm{MeV}$ of Ref.~\cite{jiang:12}, and a surface level density $A_S=0.13\unit{MeV\,fm}^{-1}$.

Once empirical and surface parametrizations are set, we use the \texttt{SROEOS} code to obtain the EOS table. 
The EOSs in the Skyrme model are obtained in the single nucleus approximation (SNA) \cite{lattimer:91, schneider:17} although extensions to accommodate multiple nuclear species have recently been proposed \cite{grams:18, furusawa:18}.  
We take the same approach discussed in our previous work and match our Skyrme-type EOSs to an EOS of 3,335 nuclei in nuclear statistical equilibrium (NSE) \cite{schneider:17}.  
A unified method to connect SNA and NSE EOSs is the subject of Refs.~\cite{gulminelli:15, raduta:19}.  
Here we follow the simple prescription to transition between SNA to NSE EOSs using a density dependent function as discussed in Sec.~VII A of Ref.~\cite{schneider:17}.  
Here, we set the transition parameters $n_\tr=10^{-3}\unit{fm}^{-3}$ and $n_\delta=0.33$, see Eqs.~(57) and (58) of Ref.~\cite{schneider:17}.
Note that the parameter $n_\tr$ is different from $n_\tr=10^{-4}\unit{fm}^{-3}$ used in Ref.~\cite{schneider:17}.  
The reason for this change is that the time to bounce in core collapse is insensitive to $n_\tr$ in the range $10^{-2}\unit{fm}^{-3} \gtrsim n_\tr \gtrsim 10^{-3}\unit{fm}^{-3}$,
while it is a function of $n_\tr$ for $n_\tr < 10^{-3}\unit{fm}^{-3}$.  
We note that setting $n_\delta \lesssim 0.5$ has little effect on CCSN simulations.  
However, larger values may have an effect since the SNA (NSE) EOS will have significant contributions at low (high) densities.

\section{Cold neutron stars}
\label{sec:NSs}

We study how variations in the empirical parameters of the EOS and of the pressure of nuclear matter at high densities affects the zero-temperature EOS and properties of cold nonrotating
beta-equilibrated NSs using the suite of EOSs discussed in Section~\ref{sec:Meta_EOS}.  
We consider each set of empirical parameter variations [see Eqs.~\eqref{eq:sets}] separately.

\subsection{Effective Mass}
\label{ssec:sM}

\begin{figure*}[htb]
\centering
\includegraphics[width=0.925\textwidth]{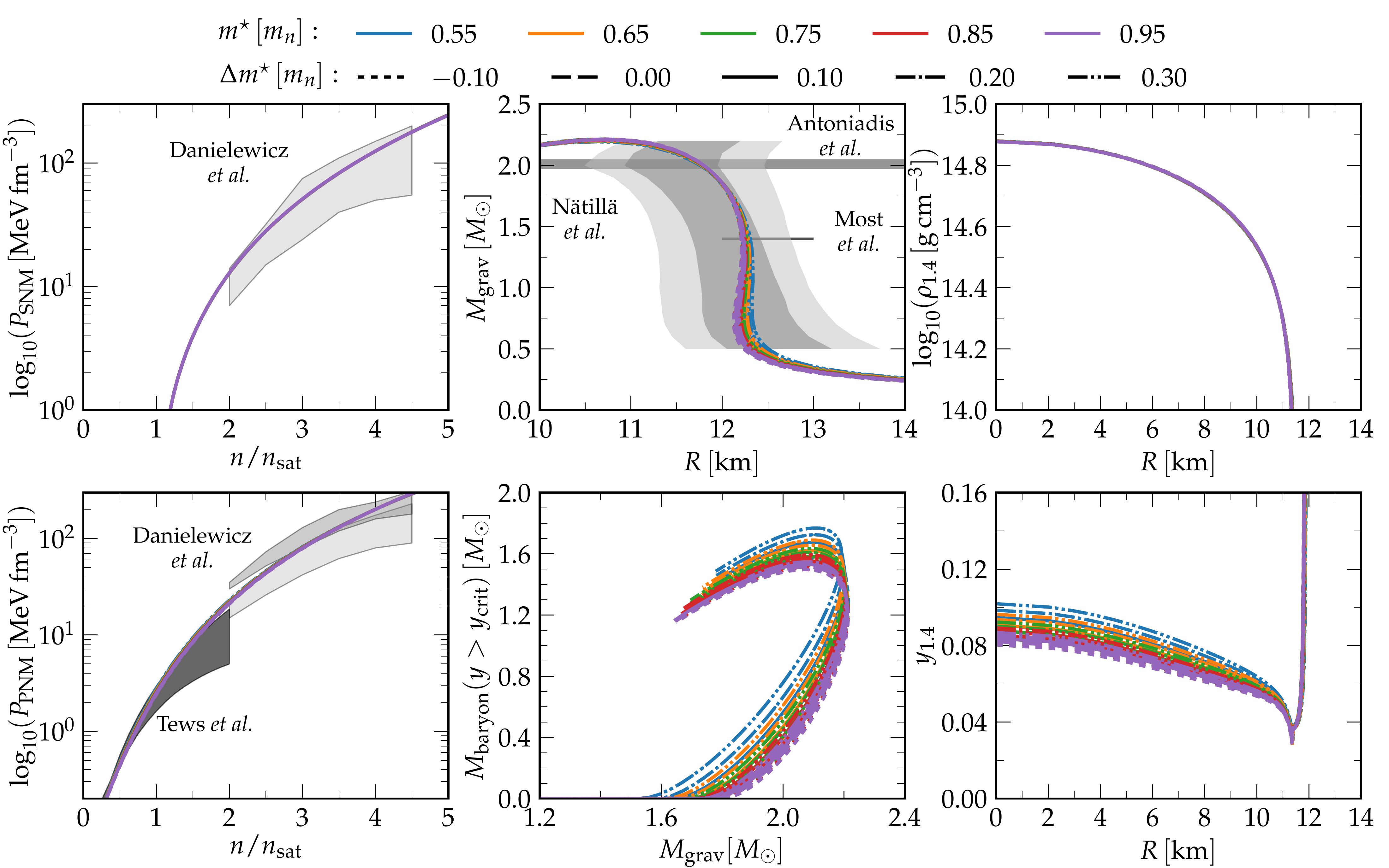}
\caption{\label{fig:NS_M} (Color online) 
Plots for variations in the effective mass $m^\star$ and effective mass splitting $\Delta m^\star$ of the pressure of SNM (top left) and of PNM (bottom left) as a function of density, of the mass-radius relations for cold beta-equilibrated NSs (top center) and the NS baryonic mass above critical proton fraction, $y_\rcrit=0.11$, as a function of the total gravitational NS mass (bottom center), and of the density (top right) and proton fraction (bottom right) as function of the radius for a canonical $1.4$-$M_\odot$ NS.
Effective masses are computed in units of the neutron vacuum mass $m_n$. 
Nuclear matter pressures are compared to results of Danielewicz~\etaln, Ref.~\cite{danielewicz:02}. 
For PNM there are two bands in Ref.~\cite{danielewicz:02} based on a strong (top band) and weak (bottom band) density dependence of the symmetry energy proposed in Ref.~\cite{prakash:88}.
PNM pressure is also compared to chiral effective field theory results of Tews~\etaln, Ref.~\cite{tews:18}. 
Mass-radius relations are compared to the mass of a NS observed by Antoniadis \etaln, Ref.~\cite{antoniadis:13}, the mass-radius relations obtained from observations of x-ray bursts by N\"attil\"a~\etaln, Ref.~\cite{nattila:16}, and the radius of a $1.4$-$M_\odot$ NS computed from the limits of tidal deformability of NSs by Most~\etaln, Ref.~\cite{most:18}.
Note that the outer $\simeq1\unit{km}$ of canonical $1.4$-$M_\odot$ NSs have densities below $10^{14}\unit{g\,cm}^{-3}$. 
All quantities plotted show only minor dependence with respect to variations in the effective mass at nuclear saturation density $m^\star$ and the neutron-proton effective mass splitting $\Delta m^\star$.
}
\end{figure*}

The Tolman-Oppenheimer-Volkoff (TOV) equations of NS structure only depend on the relationship between the pressure and energy density for the cold, beta-equilibrated EOS, $P_{\beta-\text{equil},T=0}(\epsilon_B n)$, where $n$ is the baryonic number density and $\epsilon_B$ the energy per baryon defined in Eq. \eqref{eq:Skyrme}.
Since $\epsilon_B$ and its first few derivatives are fixed at saturation density by the empirical expansion parameters, varying only the effective masses, set $s_M$ defined in Eq.~\eqref{eq:sm}, has a limited impact on $P_{\beta-\text{equil}, T=0}(\epsilon_B n)$ and one expects small variations in the nonrotating NS mass radius relation\footnote{
Due to our choice of fixing the empirical parameters of order 2 and lower in Eq.  \eqref{eq:expansion} as well as the baryonic pressures for SNM and PNM at $4n_\rsat$, 
the zero-temperature baryonic pressure, $P_B = n^2\partial\epsilon_B/\partial n$, is almost independent of $m^\star$ and $\Delta m^\star$. 
Small variations in the cold EOS for distinct choices of $m^\star$ and $\Delta m^\star$ result from how the Skyrme parameters, and, thus, the empirical parameters of order 3 and higher in Eq. \eqref{eq:expansion}, adjust to reproduce the fixed empirical parameters and the pressure at $4n_\rsat$. 
Our method contrasts with the one in Ref.~\cite{yasin:18}, where a large effect in the EOS and mass-radius relations of cold beta equilibrated NSs due to variations of the effective mass is observed. }.

The limited impact of the effective masses on the zero-temperature EOS is visible in the first column of Fig. \ref{fig:NS_M}, where we plot the zero-temperature pressures of SNM (top) and PNM (bottom) as a function of density.
No perceptible differences are seen for the EOS of SNM as the effective masses are changed. 
Meanwhile, only minor changes in the EOS of PNM occur for the different effective masses.
As in the SNM case, the EOS of PNM is, by construction, within the bounds determined from flow experiments \cite{danielewicz:02}, since we fix the pressure of PNM at four times saturation density.  
There are two bands shown for the pressure of PNM where the lower (higher) pressure band represents the pressure of PNM considering the softest (stiffest) density dependence of the PNM EOS proposed in Ref.~\cite{prakash:88}.
Our results cross the two different bands and, at the highest densities, coincide with the upper limit of the range obtained in Ref.~\cite{danielewicz:02}.
The explored range agrees with results from Ref.~\cite{tsang:18}, which compares results from flow experiments \cite{danielewicz:02} with the tidal deformability computed for the NS merger event GW170817  \cite{abbott:17a}.
We add to our comparisons the pressure of PNM obtained from chiral effective field theory (EFT) \cite{tews:18}. 
For densities up to $n\simeq1.5 n_\rsat$, the values from the Skyrme EOSs are within the constraints of chiral EFT, although they are slightly above the limits for higher densities.

In the second column of Fig. \ref{fig:NS_M}, we plot the mass-radius relations of cold beta-equilibrated NSs obtained solving the TOV equations (top) and the baryonic mass of the cold NS with proton
fraction $y$ above a critical value set to $y_\rcrit=0.11$ (bottom) as is the condition necessary for direct Urca processes to take place inside a NS \cite{brown:18}.
Because we limit our analysis to EOSs that predict a large pressure at high densities, see Tab.~\ref{tab:constraints}, all EOSs satisfy the observational constraints for the mass of PSR J0348+0432, $2.01\pm0.04\,M_\odot$ \cite{antoniadis:13}.  
A similarly large NS mass, $M=1.93\pm0.02\,M_\odot$, has been observed for PSR J1614-2230 \cite{fonseca:16}.  
Furthermore, our choices of the other empirical parameters guarantee that the mass-radius relations are within the $1\sigma$ range of ``model A'' of Ref.~\cite{nattila:16} obtained from
observations of x-ray bursts.  
The EOSs also obey the constraints for the radius of a $1.4$-$M_\odot$ NS, $12.00\unit{km}<R_{1.4}<13.45\unit{km}$, computed from the data for the NS merger observation GW170817 \cite{most:18}.  
This constraint is more stringent than obtained by others for the same event, e.~g., Ref.~\cite{soumi:18} constrain radii of NSs to be in the $8.9\unit{km} < \bar R < 13.2\unit{km}$ range while results from the LIGO and Virgo Collaborations suggest $R=11.9\pm1.4\unit{km}$ \cite{abbott:18a}.  
The constraint of Ref.~\cite{soumi:18} was computed assuming hadronic EOSs for high density matter and from inference of the dimensionless tidal deformability deduced from the GW170817 event
that suggests $\tilde{\Lambda}<800$ \cite{abbott:17a}.  
Meanwhile, the LIGO/Virgo results require that both bodies that generated the GW170817 event are NSs described by the same EOS with spins within the range observed in Galactic binary NSs and are able to produce $1.97$-$M_\odot$ NSs.  
We notice only minor differences in the mass-radius relations as a function of the effective masses, mostly in the mass range $0.5\,M_\odot \lesssim M \lesssim 1.5\,M_\odot$.

Recently, it has been shown that the cooling rate of the NS in the transient system MXB 1659-29 while in quiescence is consistent with direct Urca reactions occurring in a small fraction of the core,
$\approx0.03\,M_\odot$ \cite{brown:18}.  
Assuming hadronic matter, this is only possible if nucleons in the core are unpaired and the proton fraction exceeds a critical value $y_\rcrit$ in the range $0.11-0.15$ \cite{lattimer:91b}.
Here we set $y_\rcrit=0.11$ and compute for each NS the total baryonic mass in the core which exceeds $y_\rcrit$, $M_\rbaryon(y>y_\rcrit)$.  
We define $M_\rbaryon(y>y_\rcrit)$ as the integrated baryonic mass in regions of the star where $y\geq y_\rcrit$ excluding the crust, \ie the outer $\simeq1\unit{km}$ of the star, as densities there are to low to induce direct Urca reactions.  
If the values chosen for the empirical parameters hold, the EOS described by those parameters implies that the NS in the MXB 1659-29 system has a mass in the range $1.6$ to $1.8\,M_\odot$ as lower mass values would imply that the proton fraction in the core never reaches the critical value $y_\rcrit$ to start the direct Urca process. 
Meanwhile, NSs with larger masses would cool at a much faster rate through direct Urca processes.  
Thus, under the assumption that matter in the core of a NS is made of unpaired nucleons, combined measurements of NS masses and cooling rates may be used to improve constraints on the EOS of dense matter.

Finally, in the last column of Fig. \ref{fig:NS_M}, we compare the interior properties of a canonical $1.4$-$M_\odot$ NS for the different EOSs.  
Although there are no clear visible changes for the density as a function of NS radius, we notice that there are, as in the case of the gravitational mass with proton fraction above $y_\rcrit$, small changes in the proton fraction in the core region as a function of the nucleon effective masses.  
These variations in proton fraction in the inner core of a $1.4$-$M_\odot$ NS are inversely (directly) correlated with $m^\star$ ($\Delta m^\star$). 
However, these changes are small, and the nucleon effective masses affect the central proton fraction $y_{1.4}$ at the center of a $1.4$-$M_\odot$  NS by at most 0.02.  
Nevertheless, a clear trend is observed here: EOSs that predict smaller radii for the same mass NS also predict a larger isospin asymmetry in their cores.

\subsection{Symmetry energy and its slope}
\label{ssec:sS}

\begin{figure*}[htb]
\centering
\includegraphics[width=0.925\textwidth]{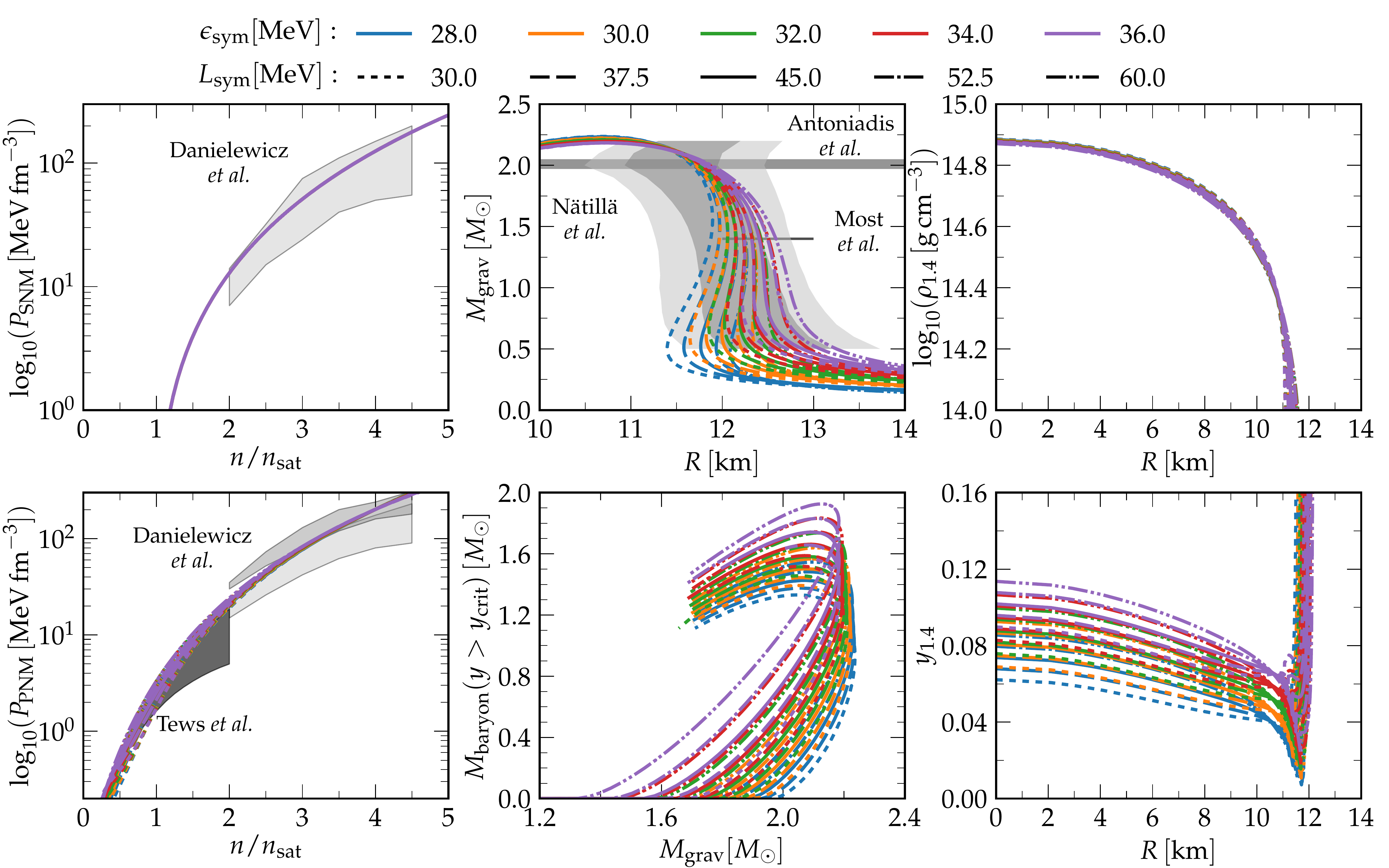}
\caption{\label{fig:NS_S} (Color online) 
Same as Fig. \ref{fig:NS_M} but for variations in the symmetry energy $\epsilon_\rsym$ and the slope of the symmetry energy $L_\rsym$. 
Both quantities are shown in units of $\unit{MeV\,baryon}^{-1}$. 
Because only the two lowest order isospin asymmetry terms are varied, the pressure of SNM (top left) is unchanged while the effects on the pressure of PNM (bottom left) are more pronounced in the region $n\lesssim2n_{\rsat}$. 
These changes impact the mass radius relationship of NSs more significantly for low mass NSs (top center). 
Meanwhile, the inner NS composition is affected even for massive NSs (bottom center). 
The difference in compositions can also be seen for canonical $1.4$-$M_\odot$ NSs, which have similar density profiles in their core (top right) but proton fractions that may differ by a factor of two (bottom right).
}
\end{figure*}

We now discuss the variation set $s_S$ [Eq.~\eqref{eq:ss}], where the symmetry energy $\epsilon_\rsym$ and its logarithmic derivative with respect to density $L_\rsym$ at saturation density are varied.

In Fig.~\ref{fig:NS_S}, we plot the pressure as a function of density and properties of cold beta-equilibrated NSs considering changes in these quantities according to Tab.~\ref{tab:constraints}.  
Because we are only modifying parameters of the symmetry energy, the pressure of SNM remains unchanged across EOSs, see the top left panel in Fig.~\ref{fig:NS_S}. 
Meanwhile, there are some variations in the pressure of PNM, as depicted by the bottom left panel of Fig.~\ref{fig:NS_S}.  
The differences between the EOSs are largest below $\simeq 2n_\rsat$ since the higher density behavior of the symmetry energy is strongly constrained by the fixed values of $P^{(4)}_\rsnm$, $P^{(4)}_\rpnm$, and $K_\rsym$ for all EOSs in the variation set $s_S$ [Eq.~\eqref{eq:ss}].
Therefore, all EOSs obey the flow constraints from Danielewicz \etal \cite{danielewicz:02} across a wide range of densities. 
In comparison, some of the $s_S$ EOSs become slightly inconsistent with the sub-saturation density chiral effective field theory constraints \cite{tews:18} at low density.

The mass-radius curve of cold beta-equilibrated NSs, the top center plot in Fig. \ref{fig:NS_S}, is most impacted by symmetry energy variations at lower NS mass.  
For NSs with mass $M\lesssim1.5\,M_\odot$, larger symmetry energies at saturation density $\epsilon_\rsym$ and symmetry energy slopes $L_\rsym$ result in larger NS radii. 
This is consistent with the results of Refs.~\cite{lattimer:01, steiner:05}, which highlight the impact of the density dependence of the symmetry energy on the NS radius.  
However, there are only minor changes in the mass-radius relationship in the region $M\gtrsim2\,M_\odot$, as these NSs reach quite high densities in their cores where the pressure is fixed by $P^{(4)}_\rsnm$ and $P^{(4)}_\rpnm$.  
Nevertheless, massive NSs with approximately the same radius have very different inner
compositions. See the bottom center panel of Fig.~\ref{fig:NS_S}.  
For the variations considered here, we observe an inverse relationship between the NS radius and the amount of matter with proton fraction larger than the critical value $y_\rcrit=0.11$, \ie the isospin asymmetry.
This is also clearly seen in the composition of the $1.4$-$M_\odot$ NS, see bottom right plot in Fig.~\ref{fig:NS_S}. 
At densities near or above $n_\rsat$, the density profile of $1.4$-$M_\odot$ NSs is similar for all EOS parametrizations that differ only  in $\epsilon_\rsym$ and $L_\rsym$, top right of Fig.~\ref{fig:NS_S}. 
However, these NS radii may differ by up to $800\unit{m}$ due to different density profiles at densities lower than $n_\rsat$.

\subsection{Incompressibility}
\label{ssec:sK}

\begin{figure*}[htb]
\centering
\includegraphics[width=0.925\textwidth]{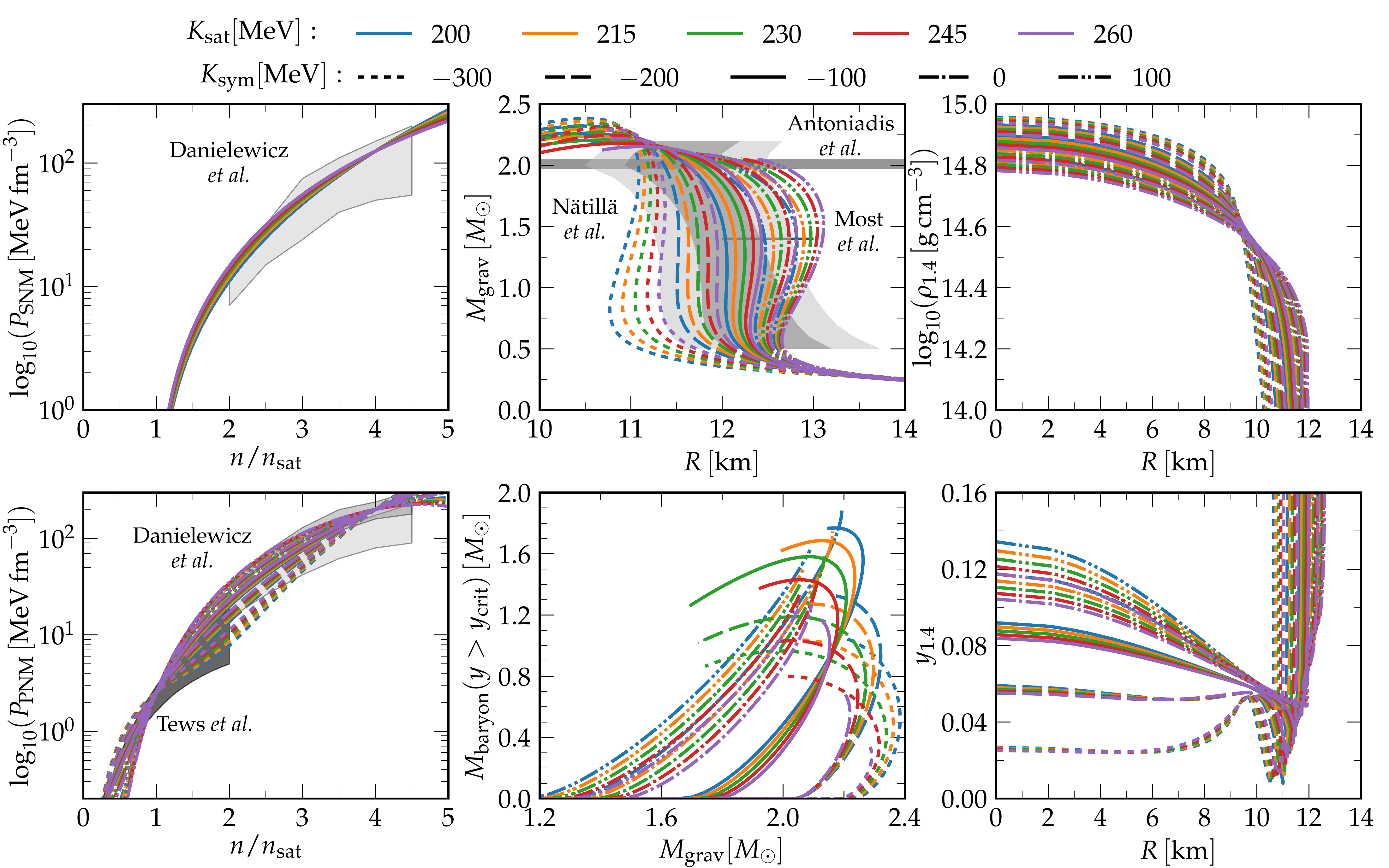}
\caption{\label{fig:NS_K} (Color online) 
Same as Fig. \ref{fig:NS_M} but for variations in the isoscalar and isovector   incompressibilities $K_\rsat$ and $K_\rsym$, respectively, measured in $\unit{MeV\,baryon}^{-1}$.  
Because of the lower uncertainty in $K_\rsat$ relative to $K_\rsym$ the variations in the pressure of SNM (top left) are smaller than those of PNM (bottom left).  
Due to the imposed constraints the pressures of both SNM and PNM match at $n=n_\rsat$ and  $n=4n_\rsat$. 
For NSs with masses lower than $\simeq2.0\,M_\odot$ there is a direct correlation between increasing incompressibility and NS radius (top center) and inverse correlation with phase space available for direct Urca processes (bottom center). 
These correlations are inverted for NSs with masses higher than $2.0\,M_\odot$.  
Canonical $1.4$-$M_\odot$ NSs are more compact for lower incompressibilities (top right) and the core proton fraction is impacted almost exclusively by the isovector incompressibility (bottom right).
}
\end{figure*}

We now consider set $s_K$ [Eq.~\eqref{eq:sk}], where we analyze variations in the isoscalar incompressibility $K_\rsat$, which is well constrained, and the isovector incompressibility $K_\rsym$, 
which is poorly known (see Section \ref{sec:Meta_EOS}).

In Fig.~\ref{fig:NS_K}, we plot the pressure of SNM (top left) and of PNM (bottom left).  
Small differences are evident in SNM for different $K_\rsat$, while the differences in PNM are substantial due to the large range of values allowed for $K_\rsym$.  
Since we keep the pressure of SNM and PNM at $n=4n_\rsat$ fixed for all EOSs, the curves
for the pressures cross at this value and at $n=n_\rsat$.
This limits the effect of both $K_\rsat$ and $K_\rsym$ at high density.

Variations in the incompressibilities cause drastic differences in the mass-radius relationships and compositions of cold NSs (see the center upper and center lower panels of Fig.~\ref{fig:NS_K}, respectively). 
There is an inverse correlation between the radius of a NS predicted by a given EOS and its isospin asymmetry, which is similar to what we see for variation sets $s_M$ and $s_S$, Secs.~\ref{ssec:sM} and \ref{ssec:sS}, respectively. 
This is particularly obvious in the rightmost panels of Fig.~\ref{fig:NS_K}, which show the internal properties of $1.4$-$M_\odot$ NSs.

We also observe different qualitative behaviors in the core composition that relate to the isovector incompressibility $K_\rsym$.
While for $K_\rsym \lesssim -200\unit{MeV}$ the proton fraction in the NS core is almost constant, for $K_\rsym \gtrsim -200\unit{MeV}$ the core asymmetry decreases with $K_\rsat$. 
Similar properties are found across NSs with the same mass but different EOSs except for the most massive ones, $M \gtrsim 2\,M_\odot$.

\subsection{Pressure at high-density}

\begin{figure*}[htb]
\centering
\includegraphics[width=0.925\textwidth]{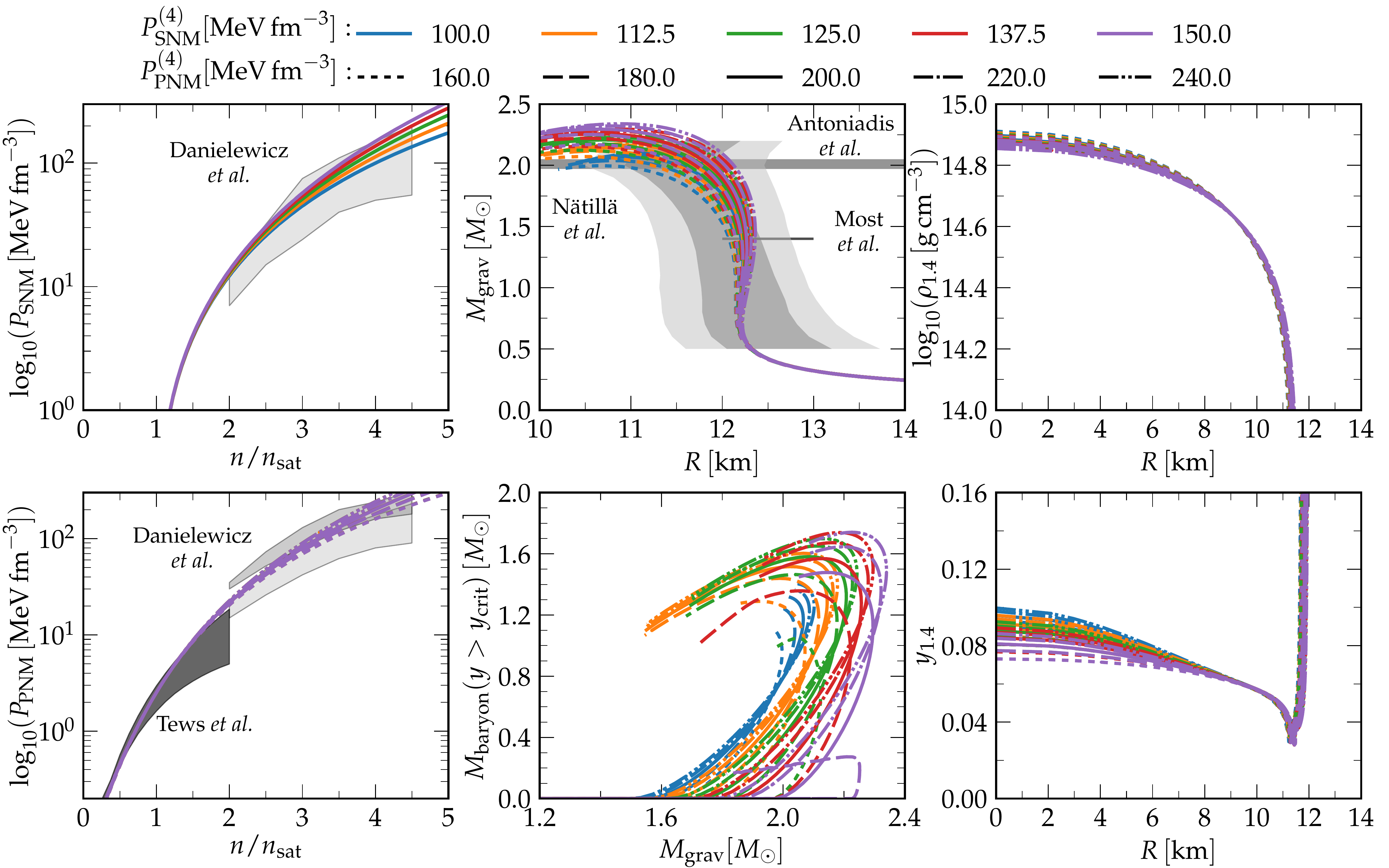}
\caption{\label{fig:NS_P} (Color online) 
Same as Fig. \ref{fig:NS_M} but for variations in the pressure of symmetric nuclear matter and pure neutron matter at $n=4n_\rsat$. 
Pressure values are given in $\unit{MeV\,fm}^{-3}$. 
In the first column we plot the pressure of SNM (top left) and PNM (bottom left). 
Higher pressures allow for higher NS masses (top center). 
Proton fraction in the core is higher for lower (higher) pressure of SNM (PNM) (bottom center).
Meanwhile, canonical $1.4$-$M_\odot$ NSs are more compact if the pressure at high densities is lower (top right). 
Again, the proton fraction in the core is higher for lower (higher) pressure of SNM (PNM) (bottom right).
}
\end{figure*}

Finally, we consider the variation set $s_P$ [Eq.~\eqref{eq:sp}], where the pressures of SNM and PNM are varied at four times nuclear saturation density, while leaving all other empirical parameters
constant. 
These variations begin to have an impact at densities $\gtrsim 2 n_\rsat$, since the saturation density properties of the EOSs are unaltered. 
This is clearly visible in the top and bottom left panels of Fig.~\ref{fig:NS_P}.

Changes in the pressure at high densities translate directly into variations in the mass-radius relationship of high-mass NSs, which probe these high densities in their cores (see the center panels of Fig. \ref{fig:NS_P}). 
Since the pressure in the NS core is somewhere between the SNM and PNM pressures, increasing either one stiffens the EOS and gives rise to a larger radius for a fixed NS mass.  
Additionally, increasing the pressure of either SNM or PNM increases the predicted maximum NS mass.

Varying these pressures also impacts the predicted lepton richness of NSs. 
In the lower left panel of Fig.~\ref{fig:NS_P}, it can be seen that the pressure of PNM is anti-correlated with the isospin asymmetry in the NS core while the pressure of SNM is correlated with the isospin asymmetry.

\section{Spherically-symmetric core collapse}
\label{sec:CCSN}

We now focus on how variations in the empirical parameters of the EOS and of the pressure of nuclear matter at high densities affects the core collapse of a massive star and its CCSN evolution.  
We will mainly investigate the impact of the EOS on neutrino emission during the postbounce phase. 
The details of neutrino emission from high-density matter in a CCSN is interesting both because these neutrinos can be directly detected from a galactic CCSN (\emph{e.g.}, \cite{scholberg:12}) and because these neutrinos can be re-absorbed in the lower density matter behind the CCSN shock and play a role in powering the explosion \cite{wilson:85}.  
Uncertainties in the nuclear EOS translate into uncertainties in predictions of CCSN neutrino fluences, which in turn introduce uncertainty in the detectability of the neutrino emission and into the CCSN mechanism itself. 
Both the explosion mechanism and detectability are sensitive to changes in the neutrino energy spectra, which we will characterize by the root-mean-square (RMS) neutrino energy, $\sqrt{\langle
\epsilon_\nu^2\rangle}$, and in the neutrino luminosities, $L_\nu$. 
Larger luminosities and RMS energies of electron neutrinos and antineutrinos result in higher predicted neutrino detection rates and more favorable conditions for explosion due to the
quadratic energy dependence of neutrino interaction cross-sections.

Specifically, we study the collapse and bounce of a $20$-$M_\odot$ progenitor star (s20WH07 of \cite{woosley:07}) in spherical symmetry using the radiation-hydrodynamics code \texttt{GR1D} \cite{oconnor:11, oconnor:15}.  
We study this progenitor star since it (1) has been studied by many other groups \cite{bruenn:13, char:15, dolence:15, melson:15b, bruenn:16, pan:16, suwa:16, summa:16, oconnor:17, bandyopadhyay:17, ott:18, just:18, glas:19, oconnor:18, oconnor:18a}, so comparisons can be readily made, (2) produces a massive PNS, (3) does not collapse into a black-hole within the first second after bounce, and (4) often
exhibits the onset of an explosion in multi-dimensional simulations \cite{bruenn:13, melson:15b, summa:16, ott:18, oconnor:18} soon after the density discontinuity from the Si/Si-O shell boundary crosses the shock radius.  
Furthermore, (5) the PNS central number density during the first second after bounce is in the range $2-3n_\rsat$. 
Since we constrain our EOS with empirical properties at saturation density and at four-times saturation density, this maximum density does not go beyond the range of densities over which the EOSs have been fit.

For each EOS table described in Section \ref{sec:Meta_EOS}, a consistent set of neutrino opacities is generated using the \texttt{NuLib} library \cite{oconnor:15}. 
We then run a core collapse simulation until $800$-$1000\,\mathrm{ms}$ after bounce. 
In the simulation, we consider electron neutrinos and electron antineutrinos separately and group the heavy flavored neutrinos and anti-neutrinos into a single composite species. 
For each species, we follow 24 logarithmically spaced neutrino energy groups running from 1 \unit{MeV} to $\simeq 269$\unit{MeV}. 
The computational grid is set to have $1\,500$ grid cells, constant cell size of $100\,\mathrm{m}$ out to a radius of $20\,\mathrm{km}$, and then geometrically increasing cell size to an outer radius of $20\,000\,\mathrm{km}$.  
We map stellar mass rest-mass density $\rho$, proton fraction $y$, and pressure $P$ from the progenitor star to \texttt{GR1D} as described in \cite{schneider:17}.

\subsection{Effective Mass}

\begin{figure*}[htb]
\centering
\includegraphics[width=0.9\textwidth]{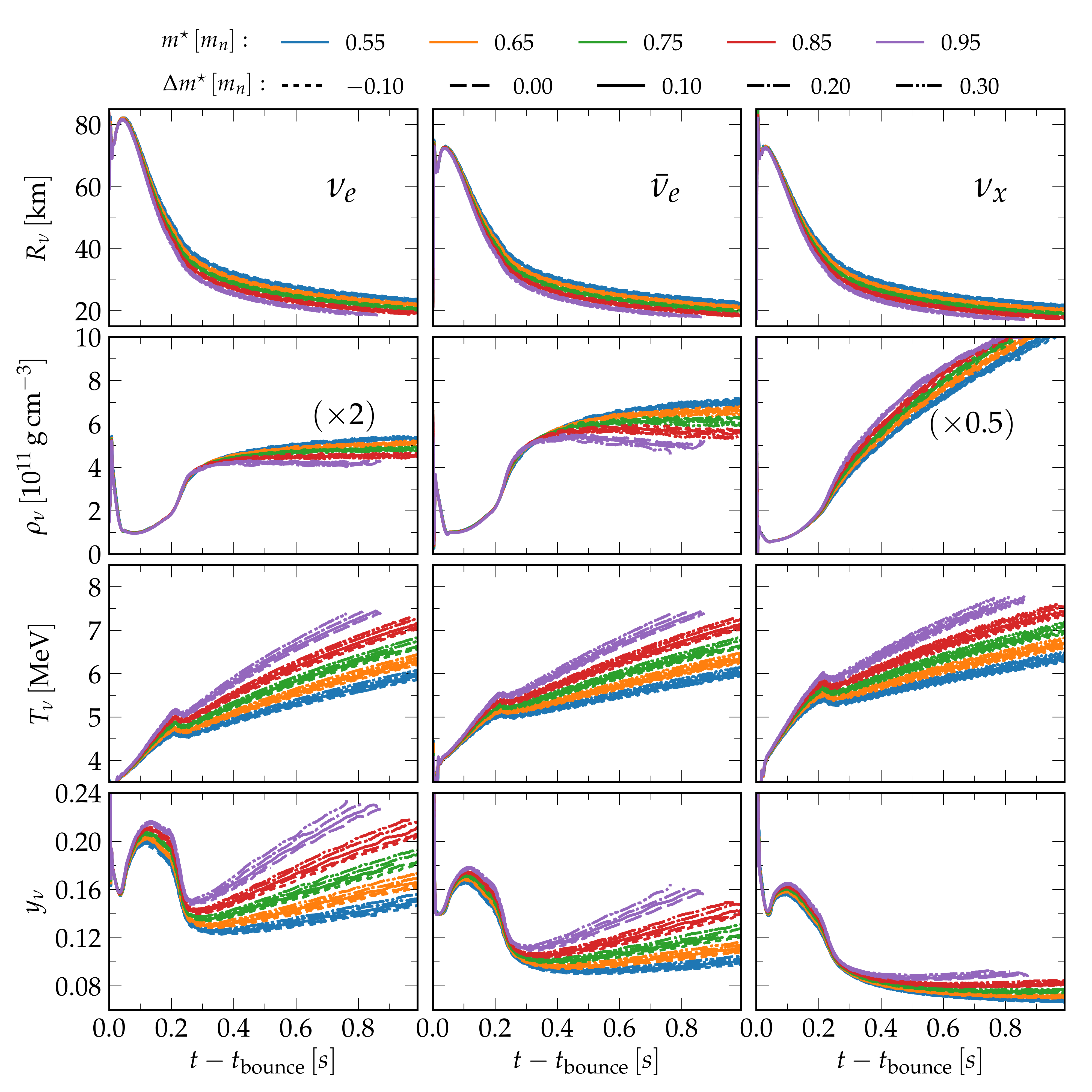}
\caption{\label{fig:nusphere_M} (Color online) 
Neutrinosphere (from top to bottom) radius, density, temperature, and proton fraction for electron neutrinos $\nu_e$ (left), electron antineutrinos $\bar\nu_e$ (center), and heavy neutrinos $\nu_x$ (right) for the spherical core collapse of the $20$-$M_\odot$ star of Woosley \& Heger \cite{woosley:07}.
We observe that increasing the EOS effective mass, $m^\star$, leads to smaller neutrinosphere radii and densities as well as higher neutrinosphere temperatures and proton fractions. 
The only exception is the $\nu_x$ neutrinosphere density which has the opposite behavior. 
Increasing the effective mass splitting, $\Delta m^\star$, has the same qualitative effect as increasing the effective mass, but to a lower order.}
\end{figure*}

\begin{figure*}[htb]
\centering
\includegraphics[width=0.9\textwidth]{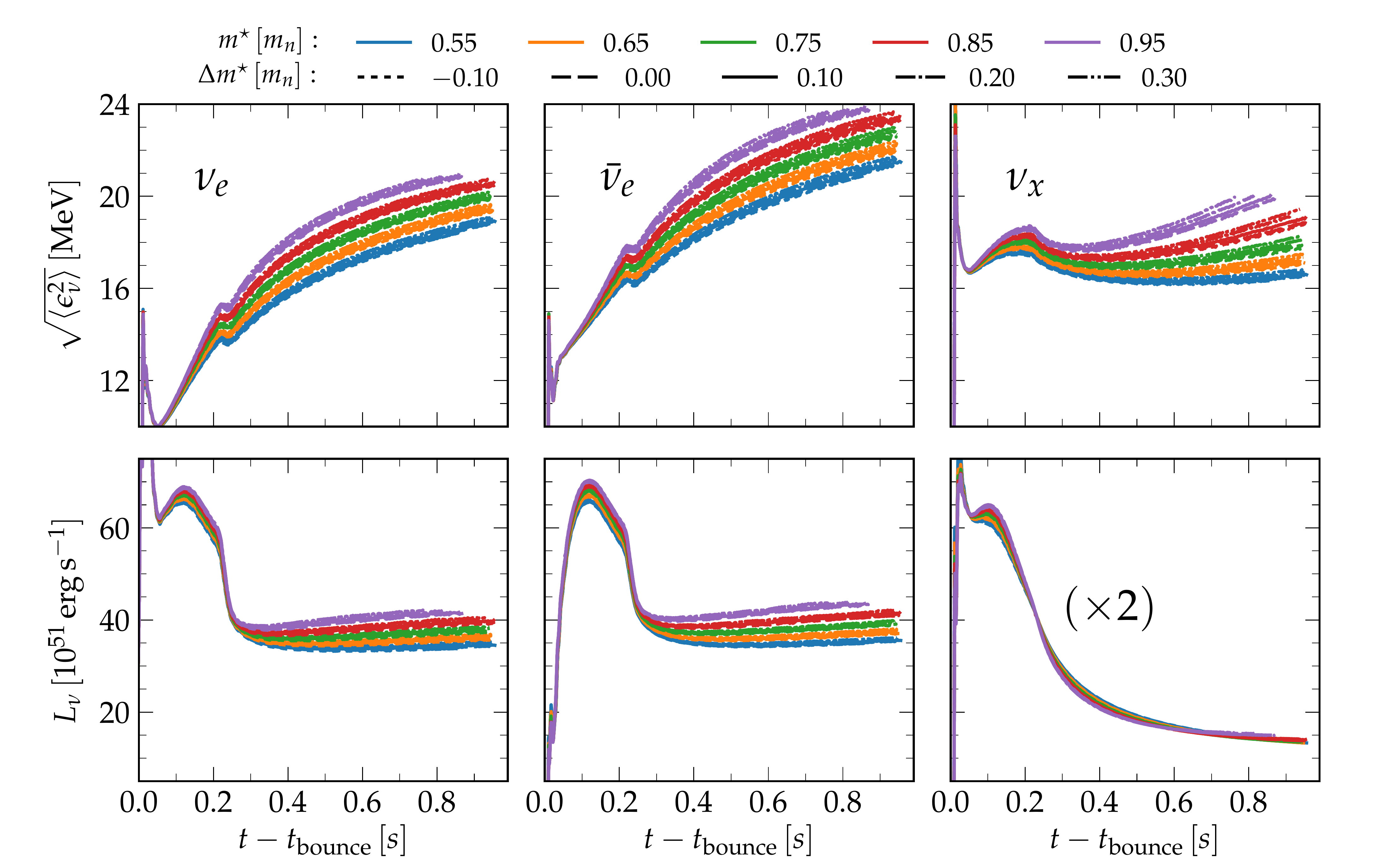}
\caption{\label{fig:nu_M} (Color online) 
Time evolution of neutrino RMS energies (top) and luminosities (bottom) for $\nu_e$ (left), $\bar\nu_e$ (center), and $\nu_x$ (right) as a function of variations in the effective masses in the EOS for the spherical core collapse of the $20$-$M_\odot$ star of Woosley \& Heger \cite{woosley:07}. 
We observe that increasing the EOS effective mass, $m^\star$, leads to higher neutrino RMS energies and luminosities. 
Increasing the EOS effective mass splitting, $\Delta m^\star$, leads to the same qualitative effect as increasing the effective mass, $m^\star$, but to a lower order. 
}
\end{figure*}

First, we consider the impact of variation set $s_M$ [Eq.~\eqref{eq:sm}] on core collapse, where the effective mass $m^\star$ and the effective mass splitting $\Delta m^\star$ are varied. 
Since the temperature enters only through the factor $m^\star T$ in the Skyrme model we use (see~Eq. \ref{eq:Skyrme}), one expects the finite temperature behavior of the EOS to be substantially impacted by changes in the effective mass. 
As shown in Sec.~\ref{sec:NSs}, varying the effective mass in our EOS fitting procedure has a negligible impact on the zero-temperature EOS and therefore a negligible impact on cold-NS structure. 
On the other hand, in CCSNe, temperatures of tens of MeV can be reached and the finite-temperature properties of the EOS may have a substantial impact.

\begin{figure}[htb]
\centering
\includegraphics[width=0.45\textwidth]{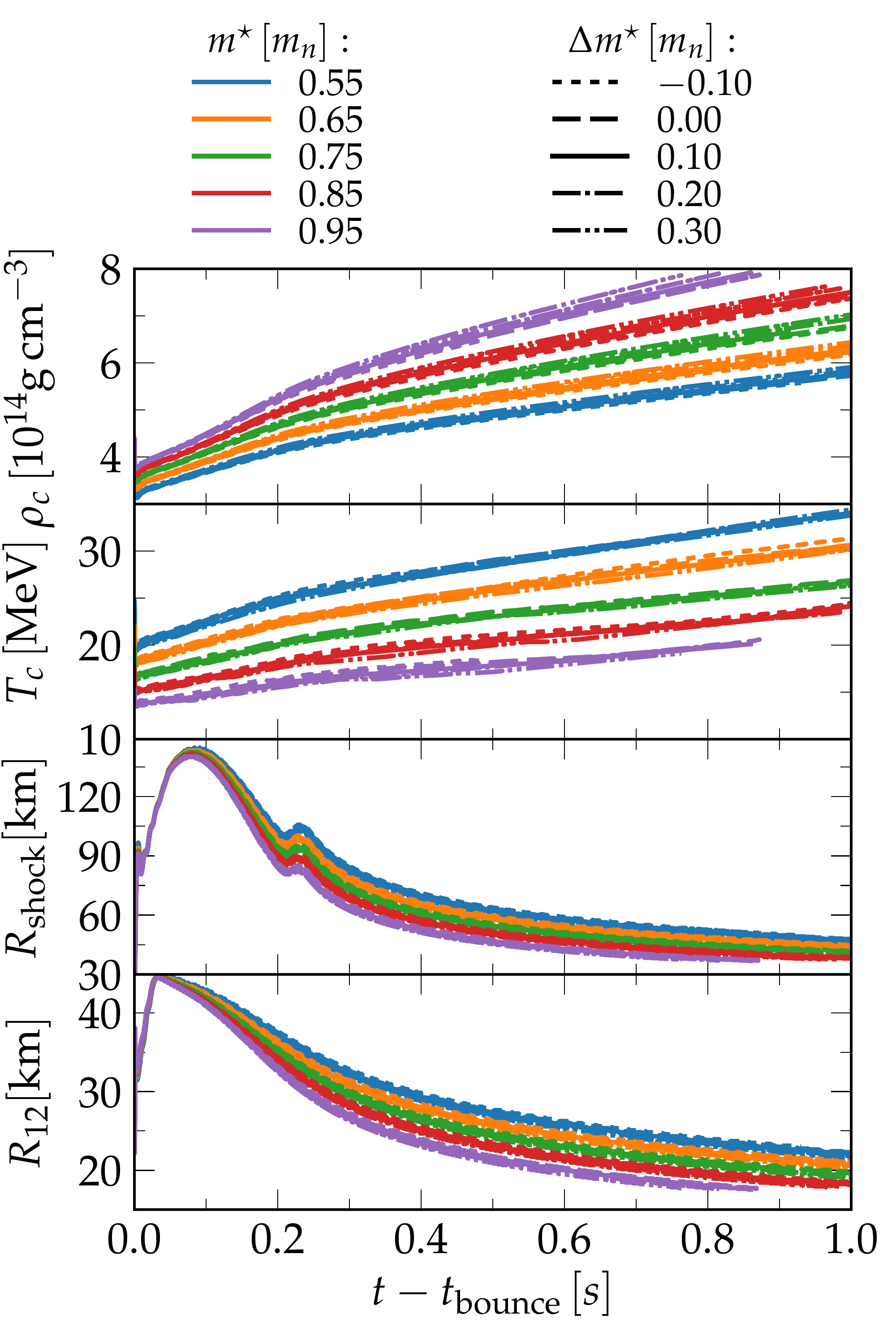}
\caption{\label{fig:GR1D_M} (Color online) 
PNS central density $\rho_c$ (top), central temperature $T_c$ (second from the top), shock radius $R_{\rm shock}$ (second from the bottom), and radius $R_{12}$ where $\rho=10^{12}\unit{g\,cm}^{-3}$ (bottom) for the spherical core collapse of the $20$-$M_\odot$ star of Woosley \& Heger \cite{woosley:07} for variations in the effective mass of SNM at saturation density, $m^\star$, and the neutron-proton effective mass splitting in the PNM limit, $\Delta m^\star$.
}
\end{figure}

\begin{figure}[htb]
\centering
\includegraphics[width=0.45\textwidth]{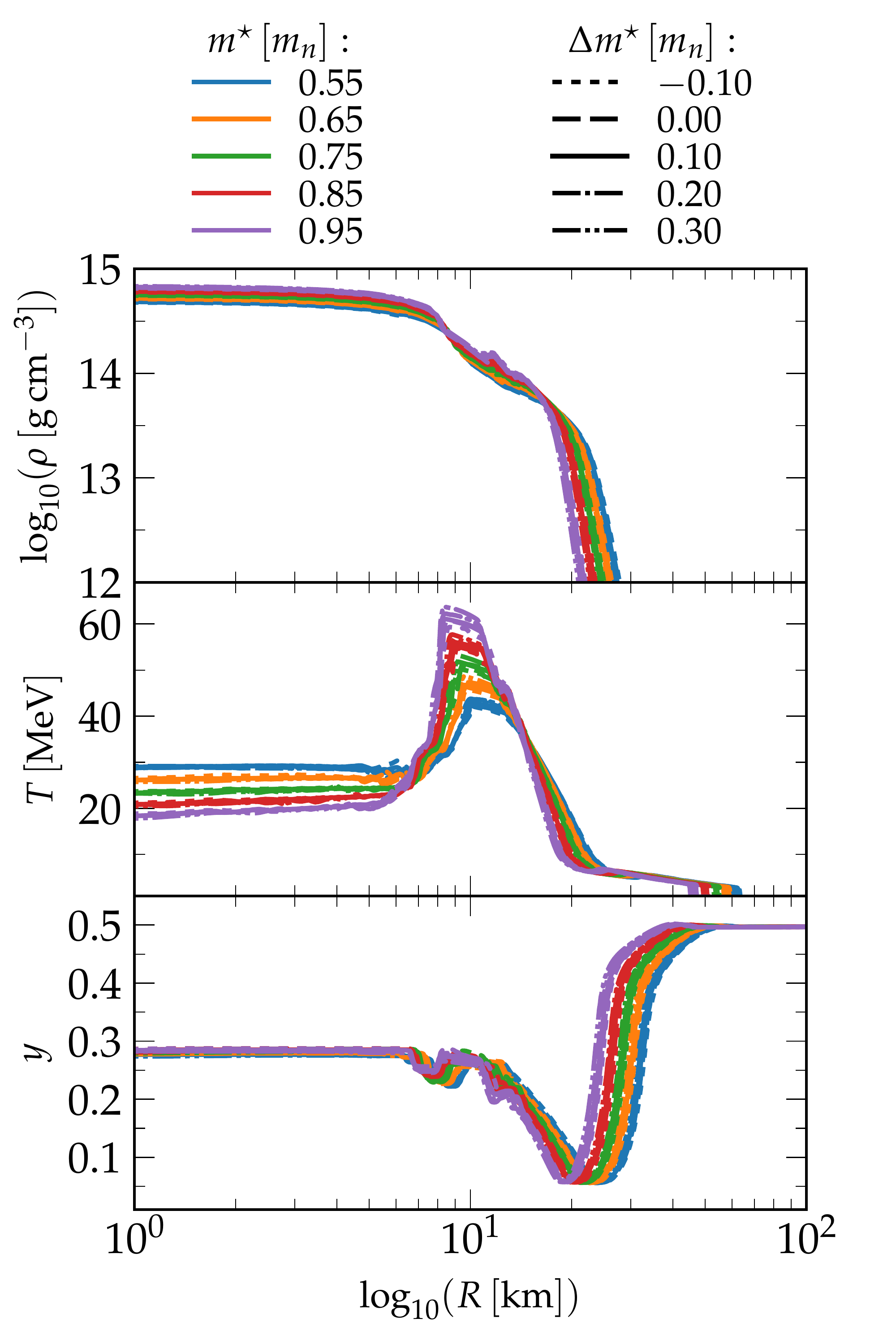}
\caption{\label{fig:500_M} (Color online) 
PNS density (top), temperature (center) and proton fraction (bottom) profiles at $500\unit{ms}$ after core bounce for the $20$-$M_\odot$ star of Woosley \& Heger 
\cite{woosley:07} for variations in the effective mass of SNM at saturation density, $m^\star$, and the neutron-proton effective mass splitting in the PNM limit, $\Delta m^\star$.
}
\end{figure}

The high-density EOS impacts the neutrino emission by changing the structure and thermodynamic state of the region from which most neutrinos are emitted, the neutrinospheres. 
The position of the neutrinosphere depends on both the energy and neutrino species (flavor, neutrino/antineutrino). Here, we consider the properties of a neutrino-energy averaged neutrinosphere, which qualitatively captures the state of the material from which the bulk of the neutrinos are emitted. 
The neutrinosphere is defined as the location where the opacity is equal to $\tau_\nu=2/3$ \cite{oconnor:13}.
Generally, before explosion, the neutrinospheres move to smaller radius, higher density, and higher temperature over time. 
The electron neutrinospheres and antineutrinospheres also stay nearly in neutrino-free beta-equilibrium since they can efficiently lose lepton number by definition.

In Fig.~\ref{fig:nusphere_M}, the influence of varying the effective mass on the neutrinosphere properties is shown. 
Increasing the SNM effective mass at saturation density, $m^\star$, increases the temperature of the neutrinosphere for all flavors and at all times.
On the other hand, increasing $m^\star$ decreases the neutrinosphere radii for all flavors.  
For electron neutrinos and antineutrinos, increasing $m^\star$ causes a decrease in the neutrinosphere density. 
Higher temperatures result in larger values of the beta-equilibrium $y_e$. 
For heavy-lepton neutrinos, increasing $m^\star$ increases the neutrinosphere density slightly. 
The impact of variations in $\Delta m^\star$ on the neutrinosphere properties is relatively small.

It is somewhat counter intuitive that the neutrinosphere temperatures increase
with the effective mass, since the effective mass enters the EOS in the
combination $m^\star T$. 
Nevertheless, it is easy to understand this behavior. First, note that deviations in the nucleon effective masses from their vacuum mass depends linearly on the density. 
Since the density of the neutrinospheres is less than a hundredth nuclear saturation density this means that $m^\star$ at the neutrinosphere is essentially the bare nucleon mass. 
Therefore, the impact of varying the effective mass on the neutrinosphere properties must be indirect. 
For small temperatures where the Sommerfeld expansion is valid, the entropy in nucleon species $t$ is given by, see Appendix \ref{app:Sommerfeld}, 
\begin{equation}\label{eq:st}
    s_t \approx \left(\frac{\pi}{\hbar}\right)^2 \frac{m^\star_t T}{(3 \pi^2 n_t)^{2/3}}.
\end{equation}
In the same approximation, the temperature dependent contribution from species $t$ to the nucleon pressure is given by $P_{\text{th},t} = Tn_t s_t/3$. 
Therefore, in so far as the density and entropy throughout the outer layers of the PNS are not impacted by changes in the effective mass, the pressure of material below the neutrinosphere
goes down with increasing effective mass since $T \propto (m^\star)^{-1}$. 
This suggests that increasing the effective mass results in more compact outer layers of the PNS. 
This is consistent with what our simulations show. As can be seen in Fig. \ref{fig:nusphere_M}, increasing the effective mass results in a smaller radius neutrinosphere which, in turn, results in a larger virial temperature for the neutrinosphere.

Variations in the neutrinosphere properties are directly imprinted in the CCSN neutrino emission itself.
In Fig.~\ref{fig:nu_M} we plot the RMS energy (top) and luminosity (bottom) of the three neutrino species considered, \ie $\nu_e$, $\bar\nu_e$, and $\nu_x = \nu_{\mu/\tau} = \bar\nu_{\mu/\tau}$.
Soon after core bounce, $t-t_{\rm bounce}\lesssim200\unit{ms}$, all EOSs predict RMS energy and luminosity of neutrinos emitted that differ only by $\lesssim5\%$ in the most extreme cases.
However, after the first $\simeq 200\unit{ms}$, neutrino energies and luminosities start to diverge.
The average RMS energy of all neutrino flavors and the luminosity for $\nu_e$ and $\bar\nu_e$ neutrinos is higher the larger the nucleon effective mass $m^\star$ at saturation density is.
Meanwhile, there is barely any change in the neutrino luminosity for the heavy-lepton neutrinos $\nu_x$ as the effective mass changes. 
Moreover, differences in neutrino properties are only affected at the $\simeq 1\%$ level by the change of the nucleon effective mass difference $\Delta m^\star$.

In Fig.~\ref{fig:nu_M} we see that the largest variation in the RMS energies occurs for the heavy-lepton neutrinos $\nu_x$ after $\simeq400\unit{ms}$ after core bounce, although the heavy lepton neutrino luminosities are barely affected. 
Nevertheless, supernova electron neutrinos and antineutrinos have a larger impact on the supernova explosion mechanism with the latter being easier to detect \cite{mirizzi:15}.
We observe that an increase in the effective mass $m^\star$ also leads to an increase in the RMS electron neutrino and antineutrino energies by about $2$ to $3\unit{MeV}$ soon after core bounce, $t-t_{\rm bounce}\gtrsim200\unit{ms}$, while luminosities increase by up to $30\%$.  
An interesting question is whether different neutrino interactions, \emph{e.g.}, inelastic neutrino-nucleus scattering, will result in the same qualitative and quantitative differences.  
We postpone an investigation of this question to future work.

An increasing effective mass increases the luminosity and average energy of electron neutrinos and antineutrinos and thereby increases the rate of neutrino heating behind the SN shock. 
Therefore, it might be expected that a higher effective mass makes conditions more favorable for shock runaway. 
Nevertheless, we find that larger effective masses result in smaller shock radii in spherically-symmetric runs. 
In Fig. \ref{fig:GR1D_M} we observe that the shock radius $R_{\rm shock}$ follows the PNS radius $R_{12}$. 
In these spherically-symmetric simulations, the impact of the reduced PNS radius on the shock overwhelms the increased neutrino heating rate when the effective mass is increased.
Nevertheless, in multi-dimensional simulations, the larger neutrino luminosities and average energies may instead lead to shock radii that expand faster for larger nucleon effective masses $m^\star$. 
This is discussed in Sec. \ref{sec:Octant}.

Besides neutrinos emitted during core collapse, we also discuss the hot PNS evolution during the first second after collapse, see Fig.~\ref{fig:GR1D_M}.
In Ref.~\cite{summa:16}, it is argued in the context of 2D simulations, that the LS220 EOS leads to fast contracting PNSs because this EOS generates compact cold beta equilibrated NSs. 
In our simulations we see that the collapse of a massive star simulated using EOSs that differ only in their effective masses predict very similar mass-radius relations for cold NSs, see Fig.~\ref{fig:NS_M}. 
Although all these EOSs produce very similar cold beta equilibrated NSs, they predict distinct behaviors for the PNSs formed in core collapse. 
In Fig.~\ref{fig:GR1D_M}, we plot the core temperature $T_c$ and density $\rho_c$ as well as shock radius $R_{\rm shock}$ and core radius $R_{12}$, the latter defined as the radius where mass density is $\rho = 10^{12} \unit{g\,cm}^{-3}$. 
There is a clear correlation between the effective mass $m^\star$ and the core density after bounce as well as how fast the PNS radius and shock contract after reaching their maximum values.
The core temperature, on the other hand, is higher (lower) the lower (higher) $m^\star$ is.

Density, temperature, and proton fraction profiles of the PNS at $500\unit{ms}$  after bounce are plotted in Fig.~\ref{fig:500_M}. 
It is clear that EOSs with higher $m^\star$ produce less thermal support since their temperatures are lower in most of the PNS interior and hot mantle, although the temperature is higher in the region where it peaks. 
Thus, we deduce the reason the LS220 EOS leads to faster contraction when compared to other EOSs is better explained by its assumptions about its effective mass, set by $m^\star=m_n$, rather than by the mass-radius relation it predicts for cold beta equilibrated NSs, which is barely affected by the effective mass.

Fig.~4 of Ref.~\cite{steiner:13} shows that the PNS radius that follows from the core collapse of a $11.2$-$M_\odot$ progenitor star simulated with the LS220 EOS contracts significantly faster than the radius of PNSs simulated with other EOSs that have $m^\star/m_n\simeq0.61-0.76$. 
However, the EOSs in that work use diverse prescriptions to compute the
EOSs at low and high densities, which makes a direct comparison between our results and their results non-trivial.
In this work, by unifying the formalism used for all EOSs, we are able to draw
stronger conclusions about the effect of each parameter of the EOS on the
core collapse, and specifically on the role of the effective mass.


\subsection{Symmetry energy and its slope}

We perform core collapse simulations using variation set $s_S$, where the symmetry energy and its slope are varied. 
We observe that for the range of variations considered for $\epsilon_\rsym$ and $L_\rsym$, the changes in the neutrino spectra and the PNS properties are rather small. 
They are of comparable in magnitude to the changes seen from varying the nucleon effective mass splitting, $\Delta m^\star$. 
Thus, for the purpose of simulations of CCSNe, these two quantities are rather well constrained and we expect that even substantial variations around the current best estimates for these two observables will not affect simulation results significantly.

It may be the case, however, that if we were to simulate these CCSNe for longer timescales, including into the cooling phase, that larger differences between EOSs could become apparent.  
We defer this, as well as CCSN simulations of different progenitors, to future work.

\subsection{Incompressibility}

\begin{figure}[htb]
\centering
\includegraphics[width=0.45\textwidth]{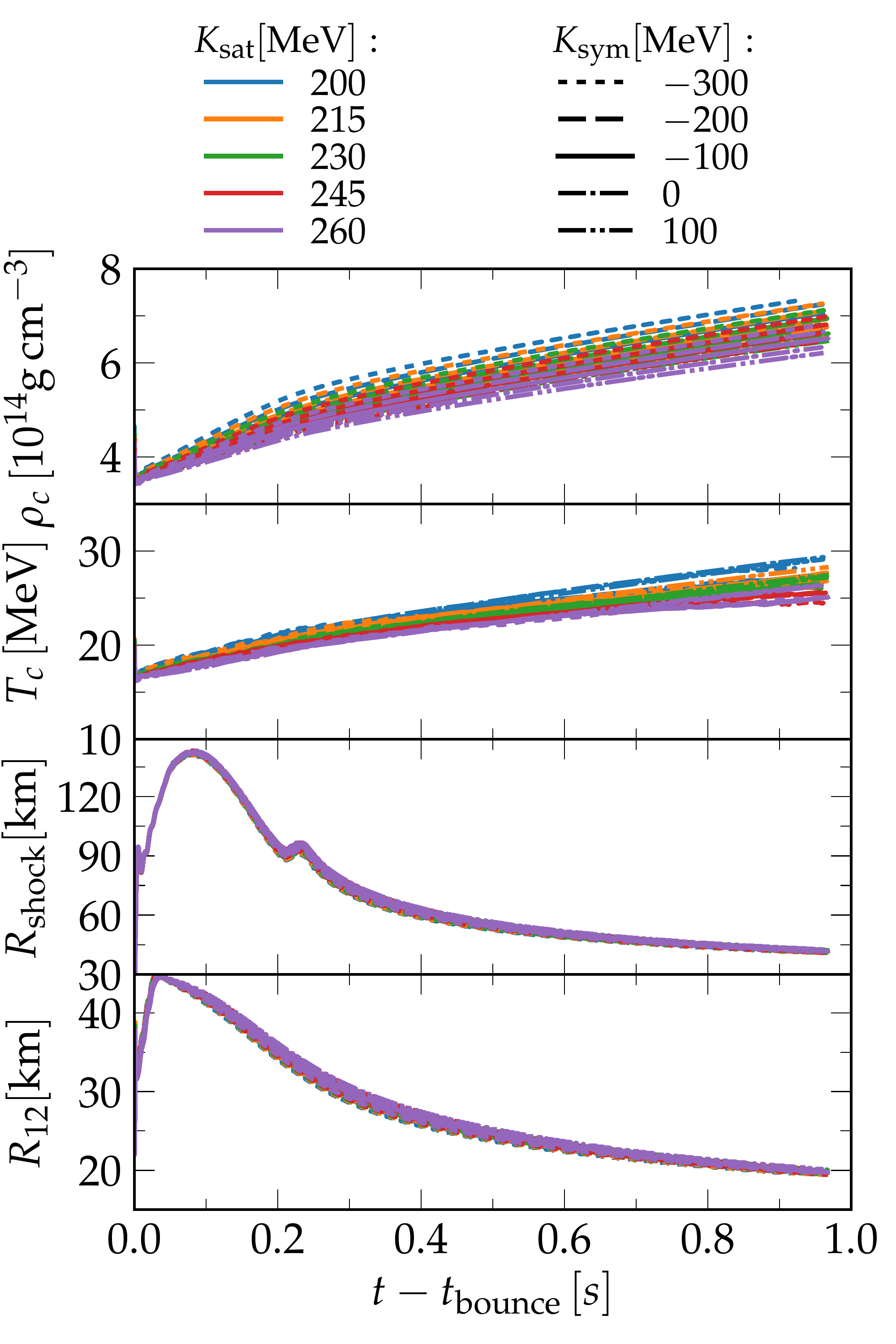}
\caption{\label{fig:GR1D_K} (Color online) 
Protoneutron star central density $\rho_c$ (top), central temperature $T_c$ (second from the top), shock radius $R_{\rm shock}$ (second from the bottom), and radius $R_{12}$ where $\rho=10^{12}\unit{g\,cm}^{-3}$ (bottom) for the spherical core collapse of the $20$-$M_\odot$ star of Woosley \& Heger \cite{woosley:07} for variations in the isoscalar and isovector incompressibilities $K_\rsat$ and $K_\rsym$, respectively.
}
\end{figure}

\begin{figure}[htb]
\centering
\includegraphics[width=0.45\textwidth]{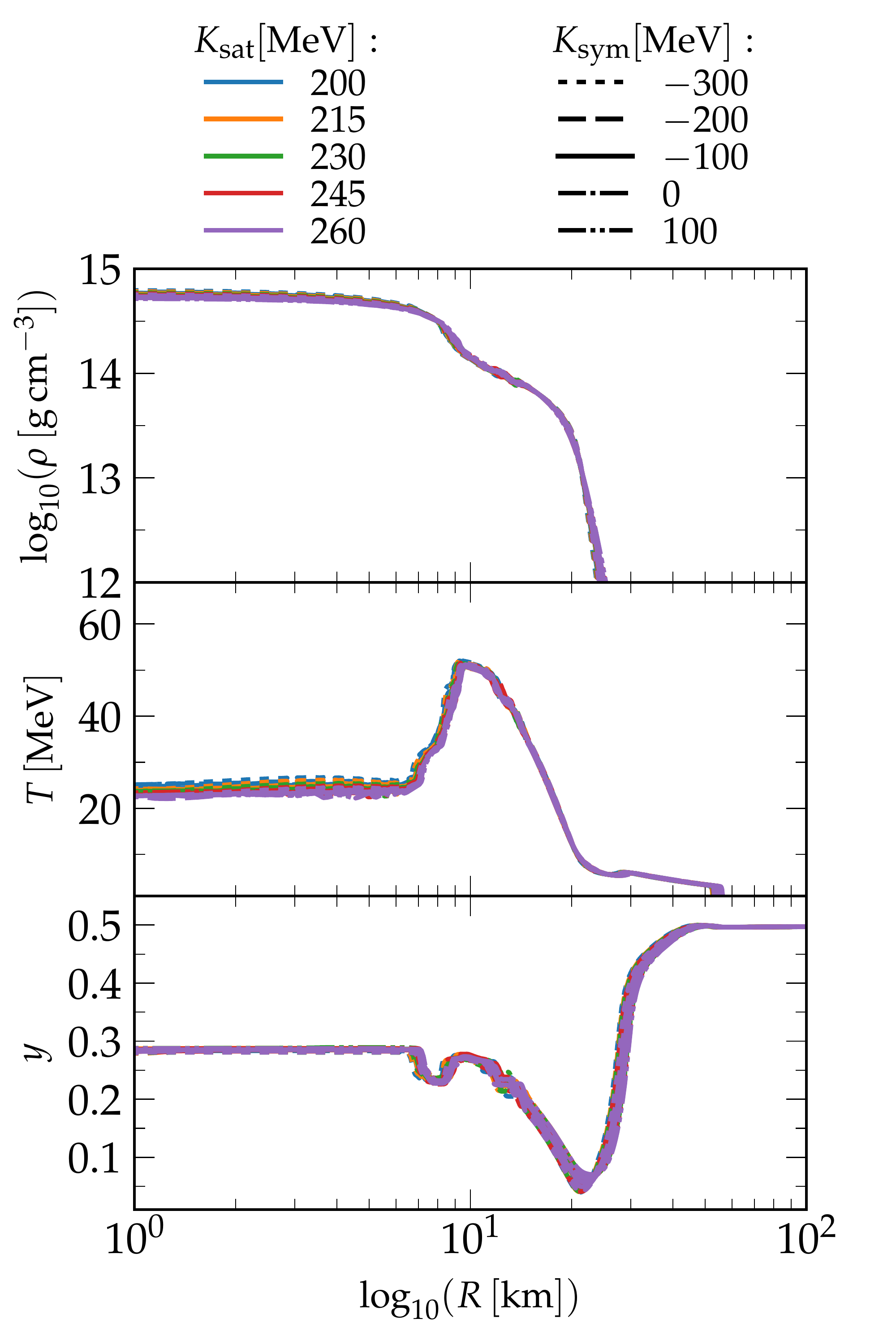}
\caption{\label{fig:500_K} (Color online) 
PNS density (top), temperature (center) and proton fraction (bottom) profiles at $500\unit{ms}$ after core bounce for the $20$-$M_\odot$ star of Woosley \& Heger \cite{woosley:07} for variations in the isoscalar and isovector incompressibilities $K_\rsat$ and $K_\rsym$, respectively.
}
\end{figure}

We now discuss effects in CCSN simulations due to changes in the incompressibility parameters $K_\rsat$ and $K_\rsym$ of the EOS.  
As in the case of variations in the symmetry energy $\epsilon_\rsym$ and its slope $L_\rsym$, the relative changes in the neutrino spectra are rather small and at most twice those observed for changes in the nucleon effective mass splitting, $\Delta m^\star$.
Nevertheless, it is likely that this is the case only for the short times we evolved the collapsing star, $t\lesssim1.0\unit{s}$.  
For longer evolutions or more massive progenitors, larger differences between the EOSs are likely.  
This may be inferred from Figs.~\ref{fig:GR1D_K} and \ref{fig:500_K}.  
The former shows the evolution of the core density, core temperature, shock radius, and PNS radius. 
The latter shows the PNS density, temperature, and proton fraction as a function of radius.  
The central density and temperature of the PNS at $\simeq1\unit{s}$ after bounce differ by  $\simeq20\%$ between the most extreme cases.  
Meanwhile, changes in the shock radius and PNS radius, are affected only in the $\simeq5\%$ range. 
At that time, the maximum PNS mantle temperature is correlated with both $K_\rsat$ and $K_\rsym$.  
On the other hand, the PNS and shock radius are anti-correlated with these quantities.  
We observe that despite the much larger error bar in $K_\rsym$ when compared to $K_\rsat$, both lead to uncertainties in PNS structure of similar magnitudes. 
We expect these differences to be amplified in multi-dimensional simulations due to the interplay between neutrino heating and hydrodynamic instabilities that can lead to shock revival \cite{muller:16}. Hence, it is important for realistic simulations that these two parameters are constrained further in the future.

\subsection{Pressure at high-density}

We also study the differences in the neutrinos spectra and in the PNS evolution during the first second of collapse for the $20$-$M_\odot$ progenitor star due to changes in the pressure of SNM and PNM at $n=4n_\rsat$, set $s_P$ in Eq.~\eqref{eq:sp}. 
As expected, by the end of our runs neither the emitted neutrinos nor the PNS properties were significantly altered by changes in the pressure at high densities. 
Except for changes of $\simeq5\%$ with respect to the baseline EOS for the density and temperature in the core near the end of the runs, none of the other quantities studied (neutrino luminosity and RMS energy, and shock and PNS radii) differed by more than 1\% during the run. 
This is due to the maximum density in the PNS still being below $2.5n_\rsat$ at $t-t_{\rm{bounce}}\simeq1\unit{s}$ and, thus, the EOSs used in all runs did not reach regions were the differences become large. 
Lower pressures at high densities cause densities (temperatures) in the core to increase faster (slower). 
As in the cases of changes in the symmetry energy $\epsilon_\rsym$ and its slope $L_\rsym$, we expect that longer evolutions will show differences for the different EOSs, as the densities reached throughout the PNS will be higher. 
Furthermore, we expect the pressure at high densities to play a significant 
role in setting the time of collapse of the PNS to a BH. 
Such a study is currently underway \cite{schneider:19}.

\section{Three-dimensional CCSN Simulations}
\label{sec:Octant}

In order to further investigate the insights gained from performing spherically-symmetric (1D) core collapse simulations discussed in Sec. \ref{sec:CCSN} for different EOSs, we perform six three-dimensional (3D) octant runs, \ie limited to one octant of the 3D cube, for the same non-rotating $20$-$M_\odot$ presupernova model s20WH07 \cite{woosley:07}. 
In our 1D simulations, we find that increasing the nucleon effective mass makes the PNS atmosphere more compact and increases the neutrino energies and luminosities. In the spherically-symmetric simulations, the impact of a reduced PNS radius overwhelmed the impact of increased neutrino heating. Hence,  larger effective masses result in smaller maximum shock radii. 
Nevertheless, spherical symmetry inhibits hydrodynamic instabilities that may be present behind the shock and these conclusions may not hold in more realistic three-dimensional simulations.
Five of the 3D runs are performed using variants of the finite temperature SLy4 EOS \cite{chabanat:98, schneider:17}. 
Additionally, we perform one run with the often used Lattimer \& Swesty EOS with $K_\rsat=220\unit{MeV}$, LS220. 
The SLy4 and LS220 EOS properties at $T=0$ are listed in Tab.~\ref{tab:SLy4}.

The variants of the SLy4 EOS are computed using the methods described in Sec.~\ref{sec:Meta_EOS} and Appendix \ref{app:linsys} by keeping all empirical quantities except the effective mass for SNM at saturation density $m^\star$ constant. 
The values used for the effective mass are $m^\star/m_n=0.6$, $0.7$, $0.8$, $0.9$, and $1.0$. 
In the discussion that follows we differentiate between the different SLy4 EOSs by adding a subscript that corresponds to the effective mass used, SLy4$_{m^\star/m_n}$. 
As in Sec. \ref{sec:CCSN}, the SLy4 EOSs as well as the LS220 EOS are connected to a low-density EOS of 3\,335 nuclei in NSE using the prescription outlined in Sec.~\ref{ssec:nonuniform}.

\begin{table}[htbp]
\caption{\label{tab:SLy4} Zero-temperature properties of the SLy4 and LS220 EOSs.}
\begin{ruledtabular}
\begin{tabular}{c D{.}{.}{4.4} D{.}{.}{4.4} l}
\multicolumn{1}{c}{Quantity}&
\multicolumn{1}{c}{SLy4} &
\multicolumn{1}{c}{LS220} &
\multicolumn{1}{c}{Units} \\
\hline
       $m^\star$ &  0.694 & 1.000  & $m_n$ \\
$\Delta m^\star$ & -0.185 & 0.000  & $m_n$ \\
\hline
$n_\rsat$        & 0.1595 & 0.1549 & $\unit{MeV}\,\rbaryon^{-1}$ \\
$\epsilon_\rsat$ & -15.97 & -16.00 & $\unit{MeV}\,\rbaryon^{-1}$ \\
\hline
$\epsilon_\rsym$ &  32.00 &  28.61 & $\unit{MeV}\,\rbaryon^{-1}$ \\
$L_\rsym$        &  45.96 &  73.81 & $\unit{MeV}\,\rbaryon^{-1}$ \\
\hline
$K_\rsat$        & 229.90 & 219.84 & $\unit{MeV}\,\rbaryon^{-1}$ \\
$K_\rsym$        &-119.70 & -24.04 & $\unit{MeV}\,\rbaryon^{-1}$ \\
\hline
$P^{(4)}_\rsnm$  & 127.12 & 107.75 & $\unit{MeV\,fm}^{-3}$ \\
$P^{(4)}_\rpnm$  & 142.15 & 162.08 & $\unit{MeV\,fm}^{-3}$ \\
\end{tabular}
\end{ruledtabular}
\end{table}


\begin{figure*}[htb!]
\centering
\includegraphics[trim={0.1cm 0.1cm 0.1cm 0.1cm},clip,width=0.925\textwidth]{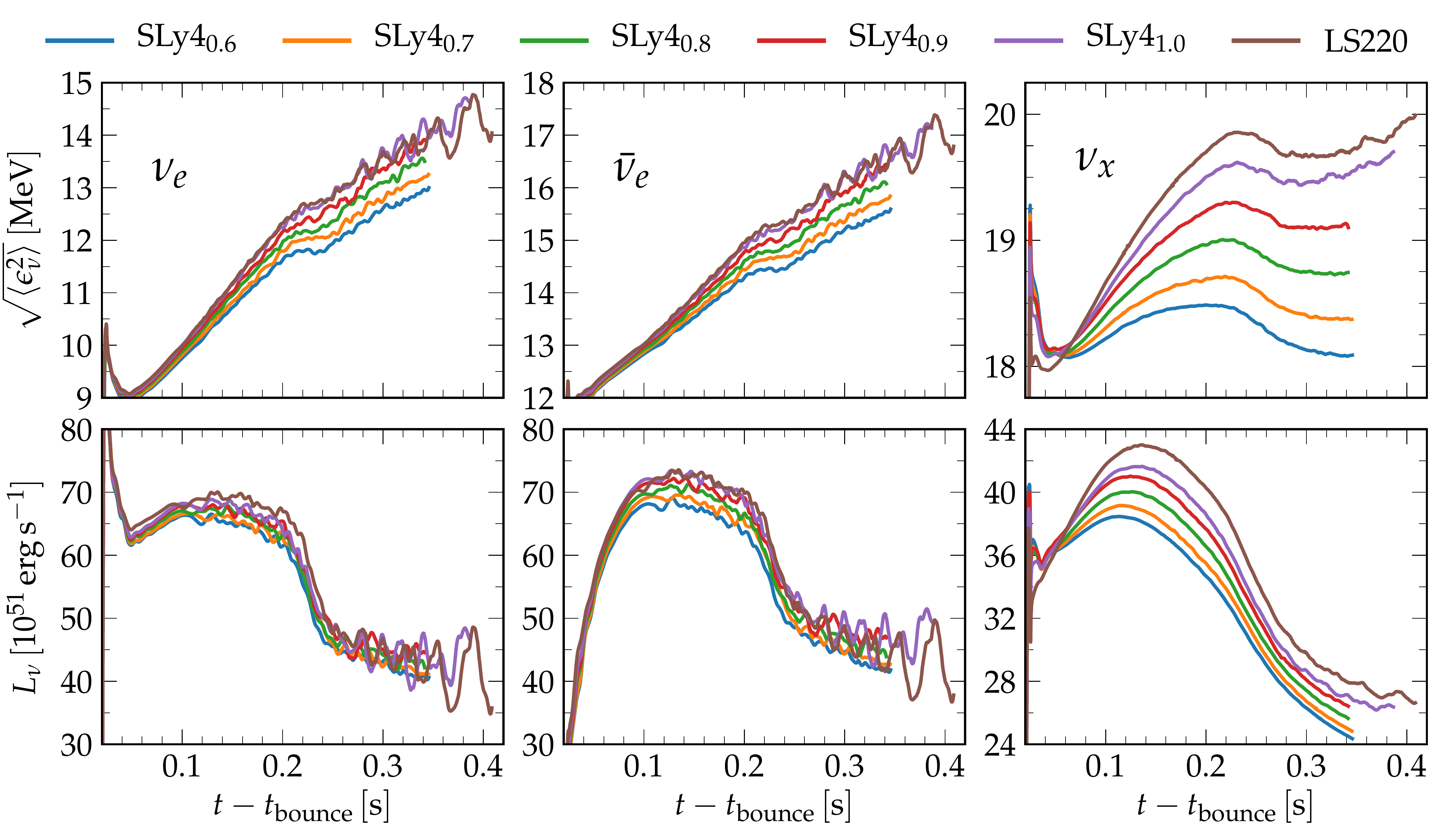}
\caption{\label{fig:nu_3D} (Color online) 
Neutrino RMS energies, $\sqrt{\langle\epsilon_\nu^2\rangle}$ (top), and luminosities, $L_\nu$ (bottom), for $\nu_e$ (left), $\bar\nu_e$ (center), and $\nu_x$ (right) for our octant runs. 
After $\simeq 100\unit{ms}$ after core bounce a clear trend appears and we observe that simulations using EOSs with higher $m^\star$ lead to higher neutrino RMS energies and neutrino luminosities. 
}
\end{figure*}

Following Sec.~\ref{sec:CCSN}, we simulate the collapse of the progenitor star using the \texttt{GR1D} code \cite{oconnor:13, oconnor:15}. 
In this phase, the neutrino reactions are considered in the exact same manner as discussed in the previous section.
Following Ref.~\cite{roberts:16}, we map the spherically-symmetric collapsing progenitor 20\unit{ms} after core bounce to a high-resolution octant 3D geometry with reflecting boundary conditions on the $xy$, $yz$, and $zx$ planes.
The remainder of the simulation is performed using the general-relativistic radiation-hydrodynamics code \texttt{Zelmani} \cite{roberts:16}, which is itself based on the \texttt{Einstein Toolkit}
\cite{loeffler:12, moesta:14a}. 
At this point we modify the neutrino transport and consider only 16 energy groups. 
As in Ref.~\cite{ott:18}, we employ the subset of neutrino opacities from Ref.~\cite{bruenn:85}, but now leave out velocity dependence and inelastic neutrino-electron scattering.

In Fig.~\ref{fig:nu_3D}, we plot the neutrino RMS energies, $\sqrt{\langle\epsilon_\nu^2\rangle}$, and luminosities, $L_\nu$, after core bounce for the three considered neutrino species. 
As in the spherically-symmetric case, both neutrino energies and luminosities, for the Skyrme-type EOSs, increase as the effective mass is increased. 
In the range of effective masses studied, differences in neutrino RMS energies are approximately $1.5\unit{MeV}$ for all neutrino species. 
We observe that neutrino energies and luminosities, especially for the heavy-lepton neutrinos $\nu_x$, computed for the LS220 EOS are higher than for the SLy4$_{1.0}$ EOS, even though both have the same effective mass for SNM at saturation density, $m^\star=m_n$. 
The reason for this is that most of the empirical parameters that differ between the two EOSs, see Tab. \ref{tab:SLy4}, shift neutrino luminosities and energies to higher values for the LS220 EOS with respect to the SLy4$_{1.0}$ EOS. 
The exception is $K_\rsym$, which slightly decreases the neutrino output for the LS220 when compared to the SLy4$_{1.0}$ EOS. 
The pressure at high densities, represented by $P^{(4)}_\rsnm$ and $P^{(4)}_\rpnm$, meanwhile, does not have a significant effect for this progenitor within the first second of core bounce.

\begin{figure*}[tb]
\centering
\includegraphics[trim={0.1cm 0.1cm 0.1cm 0.1cm},clip,width=0.925\textwidth]{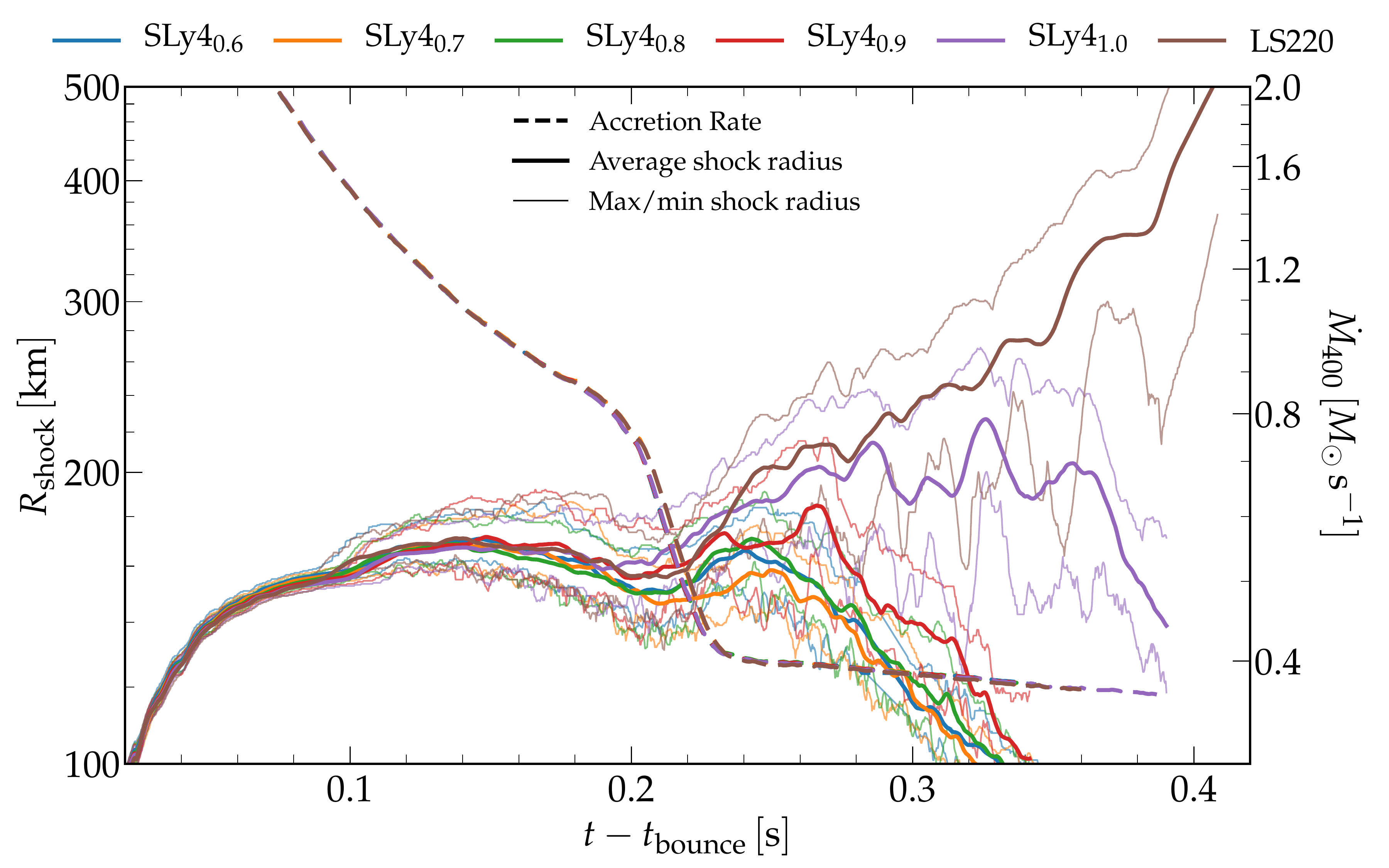}
\caption{\label{fig:shock_3D} (Color online) 
Shock radius, $R_{\rm shock}$ (solid lines, left axis), and accretion rates at $400\unit{km}$, $\dot{M}_{400}$ (dashed lines, right axis), for our octant simulations. 
Thick solid line shows the average shock radius while thin lines show the maximum and minimum shock radius. 
Accretion rates are mostly independent of the EOS and are only plotted up to the point where shock radius reaches $400\unit{km}$. 
The shock radius is very sensitive to the EOS used in the simulation, particularly after it crosses the Si/Si-O interface $\simeq220\unit{ms}$ after core bounce. 
EOSs with a higher effective mass $m^\star$ predict longer expansion of the shock radius with the LS220 EOS predicting shock runaway. 
}
\end{figure*}

Although variations of the effective mass have a similar impact on the RMS neutrino energies in the 3D simulations as they had in the 1D simulations, the
resulting shock radius evolutions differ substantially. 
The s20WH07 progenitor has a steep density and specific entropy discontinuity at the Si/Si-O shell interface. In the full 3D simulations of Ref.~\cite{ott:18} for the same progenitor star but using the SFHo EOS \cite{steiner:13}, the abrupt decrease in the ram pressure at the shock as the discontinuity is accreted results in shock runaway. 
In Fig.~\ref{fig:shock_3D}, we plot the shock radius and accretion rate for our six octant 3D simulations.
The accretion rates for all octant runs agree within 1\% or less, while the shock radius after the shock crosses the Si/Si-O is very sensitive to the EOS, with only the LS220 EOS predicting shock runaway.

In this paper, we choose not to carry out a direct comparison between our results and that of Ref.~\cite{ott:18}. We do so for a number of reasons. 
First, full 3D runs appear to more readily lead to shock runaway than octant runs \cite{roberts:16}. 
Second, when setting the initial conditions of the run we choose to preserve density $\rho$, proton fraction $y$, and \textit{pressure} $P$, while in Ref.~\cite{ott:18} chose density $\rho$, proton fraction $y$, and \textit{temperature} $T$. 
This leads to different times of core bounce and a different accretion history.
Finally, the SFHo EOS, including its low-density part, is generated using a relativistic mean-field approach and not a Skyrme model. 
Fig.~15 of Ref.~\cite{schneider:17} shows how changes in the low density EOS affect the postbounce accretion rate. 
Understanding how the difference in the low density EOS as well as in the initial conditions lead to differences in the PNS profile and CCSNe evolution is beyond the scope of the present work.

\begin{figure*}[htb]
\centering
\includegraphics[trim={0.1cm 0.1cm 0.1cm 0.1cm},clip,width=0.925\textwidth]{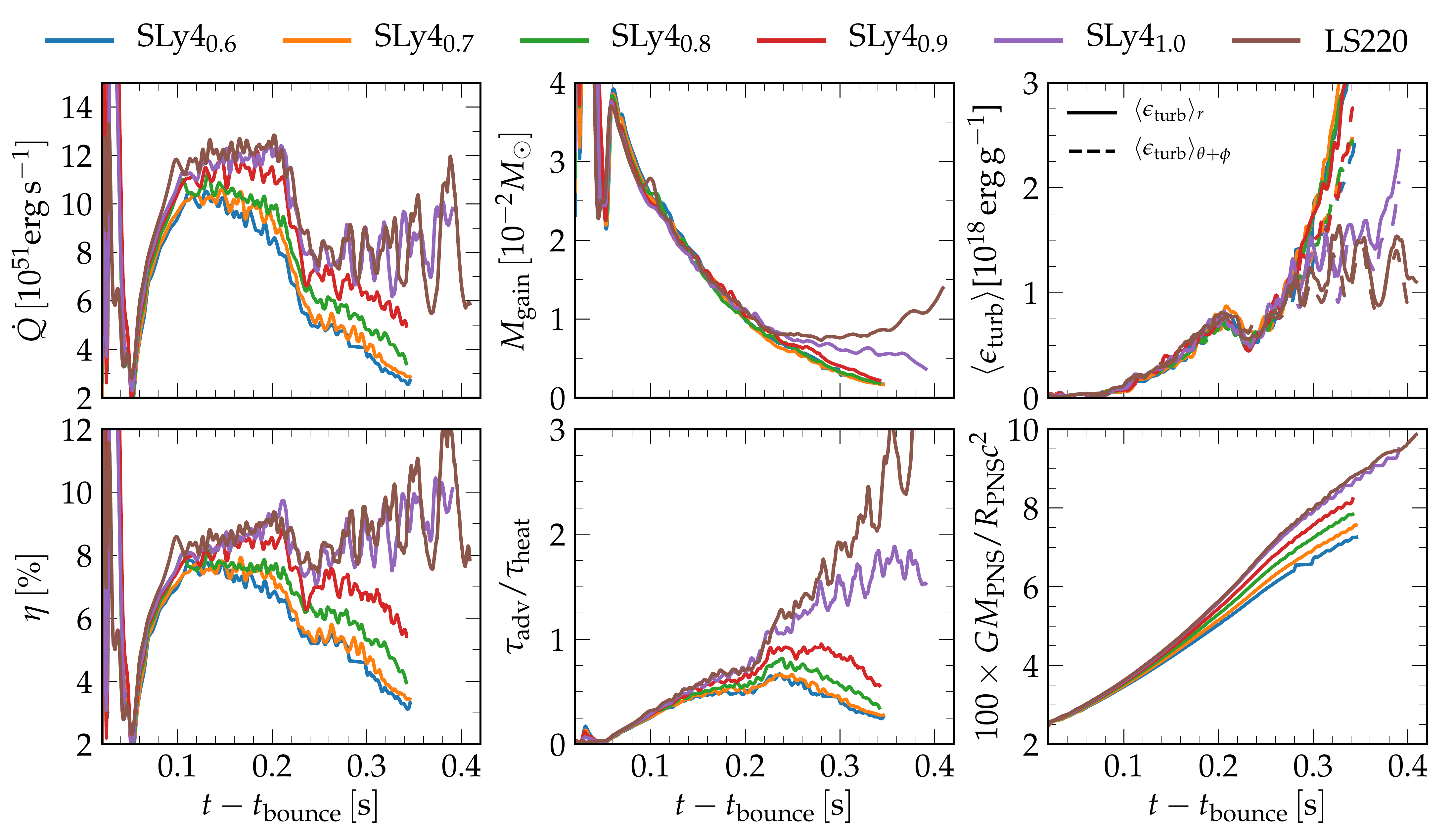}
\caption{\label{fig:panel_3D} (Color online) 
Plots of neutrino heating rate $\dot{Q}$ (top left), heating efficiency $\eta=\dot{Q}/(L_{\nu_e}+L_{\bar\nu_e})$ (bottom left), mass in the gain layer $M_{\rm gain}$ (top center), ratio $\tau_{\rm adv}/\tau_{\rm heat}$ between the mass advection $\tau_{\rm adv}$ and neutrino heating $\tau_{\rm heat}$ timescales (bottom center), turbulent energy $\langle\epsilon_{\rm turb}\rangle$ across radial and angular directions (top right), 
and PNS compactness $(G M_{\rm PNS})/(R_{\rm PNS}c^2)$ (bottom right). 
We observe a clear correlation between quantities plotted and the effective mass $m^\star$ of the EOS used in a given simulation. 
EOSs with larger $m^\star$ lead to simulations with higher neutrino heating rates, higher heating efficiency, more mass in the gain region, a larger ratio between the advection and the heating timescales, which favors shock runaway, as well as a more compact PNS. 
Total turbulent energy is anisotropic on large scales, \ie 
$\langle\epsilon_{\rm{turb}}\rangle_r \simeq \langle\epsilon_{\rm{turb}}\rangle_{\theta+\phi}$, 
nearly EOS independent up to $\simeq260\unit{ms}$, and depends on the shock radius behavior at late times, see discussion in text. 
}
\end{figure*}

With respect to the shock radius evolutions resulting from the different EOSs, we note that for the octant runs EOSs with higher effective masses for SNM at saturation density $m^\star$ generally lead to larger shock radius after bounce. 
In the LS220 run, the shock runs away approximately $350\unit{ms}$ after core bounce reaching, on average, $500\unit{km}$ by the end of the run. 
In the SLy4$_{1.0}$ run, on the other hand, the average shock radius grows up to $220\unit{km}$ at $320\unit{ms}$ after core bounce, only slightly lower than what is predicted for the LS220 EOS, but then recedes. 
Although this is opposite to the pattern seen for the shock radii in 1D runs, see Fig.~\ref{fig:GR1D_M}, this is expected in 3D simulations due to the higher neutrino luminosities and RMS energies for EOSs that have higher $m^\star$. Compare Fig.~\ref{fig:nu_M} for 1D runs and Fig. \ref{fig:nu_3D} for the 3D octant runs. 
An exception is the SLy4$_{0.6}$ EOS, whose 3D simulation predicts shock radius behavior similar to the SLy4$_{0.8}$ run and higher radii than what we observe in  the SLy4$_{0.7}$ run, despite its lower neutrino luminosities and average energies. 
This is likely due 
counteracting effects of lower neutrino production, but larger initial mass in the gain region for EOSs with lower effective masses, see Fig. \ref{fig:panel_3D}.


In Fig.~\ref{fig:panel_3D}, we present diagnostics that help us understand variations in the results for the different EOSs. 
First, higher neutrino energies and luminosities lead to higher integrated neutrino heating, heating minus cooling $\dot{Q}$, and higher heating efficiency, $\eta = \dot{Q}(L_{\nu_e}+L_{\bar{\nu}_e})^{-1}$, in the gain layer. 
Ref.~\cite{ott:18} showed that for the first $80-100\unit{ms}$ after bounce, the heating efficiency $\eta$ is almost independent of the progenitor. 
Here we observe that $\eta$ is also almost completely EOS independent early after bounce. 
However, it is clearly correlated with the effective mass $m^\star$ at later postbounce times. 
At the time when the Si/Si-O interface reaches the shock, $\eta$ is $\simeq50\%$ higher for EOSs with $m^\star=m_n$ compared to the ones with $m^\star\geq0.6m_n$.

Next, from Fig.~\ref{fig:panel_3D}, we see that the mass in the gain layer $M_{\rm gain}$ is mostly EOS independent until the Si/Si-O shell crosses the shock radius. 
After this occurs, EOSs that predict higher PNS compactness, $(G M_{\rm PNS}) / (R_{\rm PNS}c^2)$, also predict larger mass in the gain layer, another indicator of favorable conditions for shock runaway. 
The ratio between the timescales $\tau_{\rm adv}\simeq M_{\rm gain}\dot{M}^{-1}$ for material to advect through the gain layer and $\tau_{\rm heat} \simeq \vert{E_{\rm gain}}\vert{\dot{Q}^{-1}}$ for neutrino heating is another such indicator \cite{janka:01, thompson:05}. 
Following implementation details of Ref.~\cite{muller:12}, we find that two of the EOSs, LS220 and SLy4$_{1.0}$, cross the $\tau_{\rm adv} / \tau_{\rm heat} \gtrsim 1$ threshold set as a condition that favors shock runaway, while SLy4$_{0.9}$ comes very close to it. 
While the LS220 EOS results in shock runaway, none of the simulations using variants of the SLy4 EOS lead to shock runaway within $400\unit{ms}$ of core bounce. 
Not even the SLy4$_{1.0}$ EOS, despite reaching a ratio between advection and heating timescales $\tau_{\rm adv} / \tau_{\rm heat} \gtrsim 1.5$. 
As discussed in Ref.~\cite{ott:18}, $\tau_{\rm adv} / \tau_{\rm heat}$ serves more as a diagnostic of shock runaway than a condition for explosion. 
Even at times where $\tau_{\rm adv} / \tau_{\rm heat} \gtrsim 1$ for the simulations employing the SLy4$_{1.0}$ EOS, the mass in the gain layer continues to decrease and the shock stabilizes at $\langle{R_{\rm shock}}\rangle\simeq200\unit{km}$ before receding. 
In the simulation using the LS220 EOS, the mass in the gain layer stabilizes and then grows once explosion sets in.


Finally, we also plot in Fig. \ref{fig:panel_3D} the average radial and angular turbulent energies as defined in Ref.~\cite{muller:17}. 
As argued in Refs.~\cite{murphy:13,couch:15,radice:16,roberts:16} we find that the total turbulent energy is anisotropic on large scales, \ie 
$\langle\epsilon_{\rm{turb}}\rangle_r \simeq \langle\epsilon_{\rm{turb}}\rangle_{\theta+\phi}$. 
Furthermore, $\langle\epsilon_{\rm{turb}}\rangle$ is mostly EOS independent until 
$\simeq260\unit{ms}$ after core bounce, when shock behavior becomes very sensitive to the EOS. 
For simulations using the SLy4 EOSs with $m^\star\leq0.9m_n$ shock radius recedes quickly in the late stages of the run while $\langle\epsilon_{\rm turb}\rangle$ increases. 
On the other hand, using the LS220 EOS leads to saturation of $\langle\epsilon_{\rm turb}\rangle$ as its shock runs away, Fig.~\ref{fig:shock_3D}. 
The SLy4$_{1.0}$ EOS predicts a behavior for the turbulent energy density that is a mix of the predictions by the simulations using the LS220 and the other SLy4 EOSs: a momentary stabilization of $\langle\epsilon_{\rm turb}\rangle$ is achieved while $\langle{R_{\rm shock}}\rangle\simeq200\unit{km}$ followed by a fast rise as the shock radius recedes. 

\section{Conclusions}\label{sec:Conclusions}

We carried out a detailed study of the impact of variations of different 
experimentally accessible parameters of the 
nuclear matter EOS on the properties of cold beta-equilibrated neutron 
stars (NSs) and on the core collapse and postbounce evolution of 
a massive star.

Using the \texttt{SROEOS} code \cite{schneider:17}, we constructed 97
finite-temperature EOSs in which we systematically varied the
empirical parameters of the EOS based on the experimental and
theoretical constraints compiled in Refs.~\cite{margueron:18a,
  margueron:18b, danielewicz:02}.  
We then used these EOSs to compute the properties
of cold beta-equilibrated NSs and to simulate the core collapse of the
$20$-$M_\odot$ presupernova stellar model of Ref.~\cite{woosley:07}.
We carried out core-collapse supernova (CCSN) simulations using the
spherically-symmetric general-relativistic radiation-hydrodynamics
code \texttt{GR1D} \cite{oconnor:11, oconnor:10, oconnor:15}.
We carried out the
simulations to approximately and investigated the neutrino
signals and protoneutron star (PNS) evolution for each EOS.

Although the uncertainty in the effective nucleon mass at saturation 
density has a negligible impact on the properties of cold NSs in our EOS model, we find variations 
in the effective mass have a substantial impact on the postbounce evolution of our 
CCSN models. The effective nucleon mass mainly regulates the temperature dependence
of the Skyrme-type EOSs we consider, so it impacts the structure of the shock heated
material in the PNS.
Specifically, we found that the effective mass of nucleons in SNM at saturation density, $m^\star$, impacts the interior structure of the PNS, the PNS radius, the CCSN neutrino emission, and the evolution of the CCSN shock. 
Increasing the effective mass increases the average neutrino energies for all neutrino types and their total luminosity. 
This is because increasing the effective mass $m^\star$ leads to more compact PNSs with hotter neutrinospheres, although the larger effective masses result in lower PNS core temperatures. 
Recently, similar conclusions regarding the impact of the effective mass were reported from spherical-symmetric simulations of a $15$-$M_\odot$ progenitor star \cite{yasin:18}.

Variations in other parameters of the EOS, such as changes in the neutron-proton effective mass splitting in PNM, have a small impact on CCSN evolution. 
Moreover, changes in the isoscalar part of the incompressibility, $K_\rsat$, affects temperature and 
density in the core of PNSs, but has limited impact on the neutrino signal, and the outer regions of the PNS. 
Although it is more weakly experimentally constrained, varying the isospin 
incompressibility, $K_\rsym$, leads to variations in neutrinos signal and PNS
evolution of the same order of magnitude as the isoscalar incompressibility, 
$K_\rsat$. Furthermore, for the purpose of CCSN evolution, symmetry energy 
terms and the pressure at high densities, $n\gtrsim 4n_\rsat$, have even smaller 
impact on the outcome of the core collapse than changes in the incompressibility. 
Based on the spherically-symmetric simulation results, we conclude that most of the 
uncertainty introduced into simulations of core collapse evolution and its neutrino signal by uncertainties in the EOS is due to the temperature dependence of the EOS and, to a lesser
degree, due to the nuclear incompressibility.

To confirm these spherically-symmetric results, we performed six octant 3D 
simulations using the LS220 EOS and five 
variants of the SLy4 EOS where the effective mass of nucleons for SNM at 
saturation density was varied in the $m^\star=0.6-1.0\,m_n$ range. 
The runs were performed using the same set-up as the spherically-symmetric 
runs up to $20\unit{ms}$ after bounce and using the \texttt{Zelmani} code 
\cite{roberts:16} leaving out the velocity dependence and inelastic 
neutrino-electron scattering in the neutrino transport.

Among the octant runs, lower $m^\star$ causes lower neutrino average
energies and luminosities, as was the case in the
spherically-symmetric runs.  The lower neutrino energies result in
less neutrino heating of the gain layer which subsequently leads to
lower shock radii and failed explosions. Only the simulation using the
LS220 EOS ($m^\star=m_n$) shows shock runaway at $\sim350\unit{ms}$
after core bounce.  For the SLy4 EOS variants there is a strong
correlation between the shock radii and the value of $m^\star$.  For
runs employing the SLy4 EOS variant with $m^\star/m_n=1.0$,
SLy4$_{1.0}$, and $0.9$, SLy4$_{0.9}$, the average shock radius
reaches $\simeq 220\unit{km}$ and $180\unit{km}$, respectively, before
starting to recede.  For the other SLy4 EOS variants, the maximum
average shock radius is limited to $160\unit{km}$.  Analysis of our
simulations shows that the run using the SLy4$_{1.0}$ EOS reached
conditions very close to those that induce shock runaway.
Specifically, the ratio between the advection and heating time scales
is well above the limit $\tau_{\rm adv}/\tau_{\rm heat}\gtrsim1$,
usually indicative of impending shock runaway.  It is likely that the
small differences in nuclear saturation density properties between
SLy4$_{1.0}$ and LS220, which play only a secondary role in our
spherically-symmetric runs, determine that the shock runs away in the
latter simulation while it does not in the former.  We expect full 3D
simulations to more easility lead to shock runaway than the octant
simulations considered here \cite{roberts:16}.  Thus, it is likely
that for such conditions, the SLy4$_{1.0}$, and maybe even some of the
other SLy4 EOS variants with lower $m^\star$, may experience shock
runaway in full 3D.

Our octant runs may be compared to the full 3D run of Ott \etal for the same 
progenitor \cite{ott:18}. 
That run used the SFHo EOS \cite{steiner:13}, which has $m^\star=0.76m_n$. 
Nevertheless, despite the relatively low value of $m^\star$, that simulation
saw shock runaway. 
It is likely that full 3D, differences in the high and low-density EOS, and 
differences in the setup of the initial conditions all played a role in the 
outcome of that simulation.
This highlights the difficulty of comparing the role of the EOS between simulations that differ in many ways.

Understanding the effects each element of the EOS has on the outcome of a 
core collapse event is a long standing problem in nuclear and computational astrophysics. 
Using the \texttt{SROEOS} code \cite{schneider:17} we have, for the first time, 
determined in a consistent manner the pieces of the EOS that most significantly
affect core collapse dynamics and PNS evolution. 
We demonstrated that uncertainties in the temperature dependence of the EOS 
affect neutrino energies and luminosities and play an important role in determining
whether shock runaway takes place. 
We stress the need to extend our study to understand the EOS effects with 
different progenitors, full 3D simulations, and using other CCSNe simulation codes \cite{oconnor:18a} to confirm our findings.

\begin{acknowledgments}

We acknowledge helpful discussions with H.~Nagakura, I.~Tews,  C.~Constantinou, M.~Prakash, C.~J.~Horowitz, S.~Couch, and MK.L.~Warren. 
This research was funded by the National Science Foundation under award No. AST-1333520, CAREER PHY-1151197, PHY-1404569, OAC-1550514, and by the Sherman Fairchild Foundation. 
This research used resources of the National Energy Research Scientific Computing Center (NERSC), a U.S. Department of Energy Office of Science User Facility operated under Contract No. DE-AC02-05CH11231.
The authors acknowledge the Texas Advanced Computing Center (TACC) at The University of Texas at Austin for providing HPC resources that have contributed to the research results reported within this paper. 

\end{acknowledgments}

\appendix

%
%

\section{Sommerfeld Expansion}\label{app:Sommerfeld}

To compute the Sommerfeld expansion we make use of 
\begin{align}
 \lim_{T\rightarrow0}\mathcal{F}_k(\eta)=\int_0^{\eta} u^k du + 
 \frac{\pi^2}{6}T^2\left(\frac{d (u^k)}{du}\right)_{\eta} + \hdots\,.
\end{align}
Some algebra leads to
\begin{align}\label{eq:som_nt}
 n_t\simeq & \frac{2\kappa_t}{3}\tilde{\mu}_t^{3/2}
 \left[1 + \frac{\pi^2}{8}\left(\frac{T}{\tilde{\mu}_t}\right)^2\right]\,,
\end{align}
where we defined $\kappa_t=(1/2\pi^2)(2m_t^\star/\hbar^2)^{3/2}$, 
%
%
$\mu_{tF}=(\hbar^2/2m_t^\star)(3\pi^2 n_t)^{2/3}$, and $\tilde\mu_t=T\eta_t$.
We may invert Eq. \eqref{eq:som_nt} to obtain
\begin{align}
 \tilde\mu_t = \mu_{tF}
 \left[1 - \frac{\pi^2}{12}\left(\frac{T}{\mu_{tF}}\right)^2\right]\,,
\end{align}
where $\mu_{tF}=\hbar^2 k_{tF}^2/2m_t^\star$ is the Fermi chemical
potential with $k_{tF}=(3\pi^2n_t)^{1/3}$ the Fermi momentum.

A similar procedure implies that the kinetic energy density is
\begin{equation}
 \tau_t \simeq \tau_{tF}
 \left[1+\frac{5\pi^2}{12}\left(\frac{T}{\mu_{tF}}\right)^2\right]\,, 
\end{equation}
with $\tau_{tF}=\frac{3}{5}k_{tF}^{2}n_t$.
Thus, the low temperature limit of the specific entropy
\begin{align}\label{app_eq:entropy}
 s_B & = \frac{1}{T}\frac{1}{n} \sum_t\left[\frac{5}{3}
 \frac{\hbar^2\tau_t}{2m_t^\star} - T \eta_t n_t\right]\nonumber\\
     & \simeq \frac{T}{n} \frac{\pi^2}{2}\sum_t
 \left[\frac{n_t}{\mu_{tF}}\right]
 =\sum_t \frac{n_t}{n}s_t\,,
\end{align}
where $s_t$ is given in Eq.~\eqref{eq:st}.
The thermal contribution to the pressure is readily obtained from 
\begin{align}
  P_{th} & = \sum_t \left[n_t\left(\tilde\mu_t-\mu_{tF}\right) - 
  \frac{\hbar^2}{2m_t^\star}(\tau_t-\tau_{tF})\right] + Tns_B \,,
\end{align}
which reduces to $ P_{th} \simeq \tfrac{1}{3}Tns_B$. 
Expressions containing higher order terms can be found in Ref.~\cite{prakash:97}.

\section{Linear Equations}\label{app:linsys}

We present the linear equations discussed in Sec. \ref{sec:Meta_EOS} used to obtain the Skyrme parametrization given the set of EOS properties in Tab. \ref{tab:constraints}.

The $\alpha_1$ and $\alpha_2$ parameters are computed from the properties of the effective masses $m^\star(n,y)$ at two distinct points in the $n$, $y$ phase space.
We set the neutron effective mass value at $m_n^\star(n_\rsat,1/2)$ and $\Delta m^\star(n_\rsat,0)=m_n^\star(n_\rsat,0)-m_p^\star(n_\rsat,0)$ and compute the $\alpha$ parameters from the coupled equations:
\begin{align}
\label{eq:alpha1}
 (\alpha_2+\alpha_1) &=2(\beta_n^\star-\beta_n)/n_\rsat\,,\\
\label{eq:alpha2}
 \beta_\Delta&=\left(\beta_n+\alpha_1 n_\rsat\right)^{-1}
-\left(\beta_p+\alpha_2 n_\rsat\right)^{-1}.
\end{align}
Here $\beta_t=\hbar^2/2m_t$ and $\beta_\Delta=\hbar^2/2(m_n^\star-m_p^\star)$.
Eq. \eqref{eq:alpha2} reduces to $(\alpha_2-\alpha_1) n_\rsat =
(\beta_n-\beta_p)$ when $\Delta m^\star(n_\rsat,0)=0$.
We decided to compute the parameters $\alpha$ that set the effective mass of nucleons separately from the other Skyrme parameters to avoid negative effective masses at high densities and/or large isospin asymmetries.

The parameters $a_i$ and $b_i$ in Eq.~\eqref{eq:epot} are computed by solving the system of linear equations $Ax=B$ where
\begin{equation}\label{eq:linsysA}
A=
  \begin{bmatrix}
\mathfrak{a}_0 & \mathfrak{a}_0 &
\mathfrak{a}_1 & \mathfrak{a}_1 &
\mathfrak{a}_2 & \mathfrak{a}_2 &
\mathfrak{a}_3 & \mathfrak{a}_3 \\[5pt]
\mathfrak{a}_0' & \mathfrak{a}_0' &
\mathfrak{a}_1' & \mathfrak{a}_1' &
\mathfrak{a}_2' & \mathfrak{a}_2' &
\mathfrak{a}_3' & \mathfrak{a}_3' \\[5pt]
0 & -\mathfrak{a}_0 & 
0 & -\mathfrak{a}_1 & 
0 & -\mathfrak{a}_2 & 
0 & -\mathfrak{a}_3 \\[5pt]
0 & -3\mathfrak{a}_0' & 
0 & -3\mathfrak{a}_1' & 
0 & -3\mathfrak{a}_2' & 
0 & -3\mathfrak{a}_3' \\[5pt]
\mathfrak{a}_0'' & \mathfrak{a}_0'' &
\mathfrak{a}_1'' & \mathfrak{a}_1'' &
\mathfrak{a}_2'' & \mathfrak{a}_2'' &
\mathfrak{a}_3'' & \mathfrak{a}_3'' \\[5pt]
0 & -\mathfrak{a}_0'' & 
0 & -\mathfrak{a}_1'' & 
0 & -\mathfrak{a}_2'' & 
0 & -\mathfrak{a}_3'' \\[5pt]
\mathfrak{b}_0 & \mathfrak{b}_0 &
\mathfrak{b}_1 & \mathfrak{b}_1 &
\mathfrak{b}_2 & \mathfrak{b}_2 &
\mathfrak{b}_3 & \mathfrak{b}_3 \\[5pt]
0 & -\mathfrak{b}_0 & 
0 & -\mathfrak{b}_1 & 
0 & -\mathfrak{b}_2 & 
0 & -\mathfrak{b}_3 \\[5pt]
  \end{bmatrix}\,,
\end{equation}
where we defined 
\begin{align}
 \mathfrak{a}_i   &= n_\rsat^{\delta_i}\\
 \mathfrak{a}_i'  &= \delta_i n_\rsat^{\delta_i}\\
 \mathfrak{a}_i'' &= 9\delta_i (\delta_i-1) n_\rsat^{\delta_i}\\
 \mathfrak{b}_i   &= \delta_i \left(4 n_\rsat\right)^{\delta_i}\,,
\end{align}
$x=(a_0,b_0,a_1,b_1,a_2,b_2,a_3,b_3)^T$, and
\begin{gather}\label{eq:linsysb}
  B=
  \begin{bmatrix}
\epsilon_\rsat-\epsilon_\rkin(n_\rsat,0.5) \\[5pt]
{n_\rsat^{-1}}P_\rkin(n_\rsat,0.5) \\[5pt]
\epsilon_\rsym-\epsilon_{\rsym,\rkin}(n_\rsat,0.5) \\[5pt]
L_\rsym-L_{\rsym,\rkin}(n_\rsat,0.5) \\[5pt]
K_\rsat-K_\rkin(n_\rsat,0.5) \\[5pt]
K_\rsym-K_{\rsym,\rkin}(n_\rsat,0.5) \\[5pt]
P^{(4)}_{\rsnm}-P_\rkin(4n_\rsat,0.5)\\[5pt]
P^{(4)}_{\rpnm}-P_\rkin(4n_\rsat,0)
  \end{bmatrix}\,.
\end{gather}
In Eq.~\eqref{eq:linsysb} $n_\rsat$, $\epsilon_\rsat$, $\epsilon_\rsym$,
$K_\rsat$, $K_\rsym$, $L_\rsym$, $P^{(4)}_{\rsnm}$, and $P^{(4)}_{\rpnm}$, are, respectively, the nuclear saturation density, energy at saturation, symmetry energy at nuclear saturation density, isoscalar incompressibility, isovector incompressibility, the slope of the symmetry energy, and the pressures of SNM and PNM at $4n_\rsat$.
Furthermore, $\epsilon_\rkin(n,y)$ is the kinetic energy term of the specific energy and was defined in Eq. \eqref{eq:ekin} while
\begin{align}
P_\rkin(n,y) &= n^2\left.\frac{\partial\epsilon_\rkin(n',y)}
{\partial n'} \right|_{n}\,, \\
K_\rkin(n,y) &= 9n^2
\left.\frac{\partial^2\epsilon_{\rkin}(n',y)}{\partial n'^2}\right|_{n}\,, \\
K_{\rsym,\rkin}(n,y) &= 9n^2
\left.\frac{\partial^4\epsilon_{\rkin}(n',y')}{\partial y'^2n'^2}\right|_{n,y}\,, \\
\epsilon_{\rsym,\rkin}(n,y) &= \frac{1}{8}\left.
\frac{\partial^2\epsilon_\rkin(n,y')} {\partial y'^2} \right|_{n,y}\,, \\
L_{\rsym,\rkin}(n,y) &=
\frac{3}{8}n\left.\frac{\partial^3\epsilon_\rkin(n',y')}
{\partial y'^2\partial n'} \right|_{n,y}\,.
\end{align}

\bibliography{Meta_EOS}

\end{document}